\documentstyle[12pt]{article}
      
\begin{document}\begin{flushright}\thispagestyle{empty}
OUT--4102--86\\
DTP/00/24 \\    
hep-ph/0007152\\
\end{flushright}
\begin{center}{
                                                    \Large\bf
Renormalons and multiloop estimates in scalar        \\[4pt]
correlators, Higgs decay and quark-mass sum rules
                                                    }\vglue 10mm{\large{\bf
D.~J.~Broadhurst$^{a,1)}$,
A.~L.~Kataev$^{b,2)}$ and
C.~J.~Maxwell$^{c,3)}$                              }}

$^{a)}$ Physics Department, Open University, Milton Keynes MK7 6AA, UK

$^{b)}$ 
Institute for Nuclear Research of the Academy of Sciences of Russia,\\ 
117312 Moscow, Russia

$^{c)}$ Centre for Particle Theory, University of Durham, Durham, DH1 3LE, UK
\end{center}
{\bf Abstract.}
The single renormalon-chain contribution to the correlator of
scalar currents in QCD is calculated in the $\overline{\rm MS}$-scheme 
in the limit of a  large number of fermions, $N_f$. At $n$-loop order
we find that in the $\overline{\rm MS}$-scheme the factorial 
growth of the perturbative coefficients due to renormalons 
takes over almost immediately in the euclidean region. The essential 
differences between the large-order growth of perturbative coefficients
in the present scalar case, and in the previously-studied vector case are
analysed.
 In the timelike region a stabilization 
of the corresponding perturbative series for the imaginary part, 
with $n$-loop behaviour  
$S_n/[\log(s/\Lambda^2)]^{n-1}$, where $S_n$ is essentially constant for
$n\le{6}$, is observed. Only for $n\ge{7}$ does one discern the factorial
growth and alternations of sign. 
We use the new all-orders  results 
to scrutinize the performance of multiloop estimates , using
a large-$\beta_0=(11N_c-2N_f)/12$ approximation, the so-called 
``naive nonabelianization'' procedure, and within the effective charges 
approach.
The asymptotic behaviour of perturbative coefficients, in both
large-$N_f$ and large-$N_c$ limits, 
is analysed both in the spacelike and timelike regions.
A contour-improved 
resummation technique in the time-like region is developed. 
Some subtleties connected with scheme-dependence
are analysed , and illustrated using results in
 the $\overline{\rm MS}$ and $V$-schemes. 
The  all-orders series under investigation are summed up with the help 
of the Borel resummation method. The results obtained are relevant to 
the analysis of the theoretical uncertainties in the 4-loop extractions 
of the running and invariant $s$-quark masses from QCD sum rules, and in calculations
of the Higgs boson decay width into a quark-antiquark pair.

PACS : 11.15.Bt; 11.15.Me; 12.38.Bx; 12.38.Cy

Key words: renormalons, QCD perturbative series, 
resummation of the effects of analytical continuation.

\vfill\footnoterule\noindent
$^1$) D.Broadhurst@open.ac.uk;
http://physics.open.ac.uk/$\;\widetilde{}$dbroadhu\\
$^2$) Kataev@ms2.inr.ac.ru\\
$^3$) C.J.Maxwell@durham.ac.uk.

\newpage\setcounter{page}{1}
\newcommand{\ep}{\varepsilon}
\newcommand{\de}{\delta}

\section{Introduction}

Thanks to fine work by Chetyrkin~\cite{KC}, we have
information about the imaginary part of the correlator of a pair
of scalar currents at the 4-loop level of QCD,
i.e.~to the same order in perturbation theory as for the vector channel
\cite{GKLR}.
The role of the vector correlator, in the annihilation of an
electron-positron pair into hadrons, has been much studied.
Here we focus on the scalar channel, which is even more intriguing.
There are at least 6 potent reasons for studying scalar correlators.

\noindent{\bf 1.~Quark masses:}\quad
Consider the scalar divergence,
$i(m_a-m_b)\overline\psi_a\psi_b$,
of a flavour-changing vector current.
In perturbative QCD, its correlator is known exactly~\cite{DJB}
to two-loop order, where it involves trilogarithms of two variables:
$m_a^2/Q^2$ and $m_b^2/Q^2$, at euclidean momentum $Q^2$.
At 3 loops, it is was possible to obtain the first two~
\cite{Gor1,Gor2,Gor3,3M2} terms
of the expansion in $1/Q^2$; 
at 4 loops, only the imaginary part is known,
and then only to leading order in $1/Q^2$. The coefficients of
quark and gluon condensates are known, to lesser accuracy \cite{ST}.
Combining this information with experimental data on $K_0^*$
mesons, coupling to the scalar form factor of semi-leptonic
$K_{l3}$ decay, and with further theoretical input
from chiral perturbation theory,
the scale-dependent strange-quark mass of the modified
minimal subtraction ($\overline{\rm MS}$)
scheme has been determined~\cite{CPS} 
at a scale
of $\mu=1$~GeV, from whence it may be evolved to higher scales,
using the anomalous quark-mass dimension
$\gamma_m(\alpha_s)=d\log m(\mu^2)/d\log(\mu^2)$.

\noindent{\bf 2.~Higgs decay:}\quad
At the opposite extreme of very high energy, 4-loop perturbative
analysis of the scalar channel yields radiative corrections~\cite{KC}
to the decay of the Higgs boson of the standard model into quark-antiquark
pairs, with a coupling to each flavour proportional to the mass of the quark
(the 3-loop corrections are known from the results of 
Refs.\cite{Gor2,Gor3,3M2,ChKSt})).\\

\noindent{\bf 3.~Renormalization group:}\quad
As is clear from these two important phenomenological investigations,
mass renormalization is of the essence in the scalar channel.
In the vector channel, we may ignore quark-mass effects at high energy;
in the scalar channel they abide, since the
only form of scalar current that has meaning
is $\overline\psi_a M_{a,b}\psi_b$, where $M$
is a mass matrix. Hence the vertex
renormalization of the scalar current $\overline\psi_a\psi_b$
is precisely the inverse of mass renormalization.
It follows that the anomalous quark-mass dimension
$\gamma_m(\alpha_s)$ is ever present in the renormalization-group
equations for correlators of scalar currents, while in
the vector case it is inactive at very high energy,
provided the order $O(m^2/Q^2)$  corrections 
are neglected.

Thanks to the dedicated labour and great ingenuity of colleagues,
we have been provided with the 4-loop anomalous quark-mass
dimension~\cite{4LM1,4LM2} and the 4-loop beta function~\cite{4LB} of QCD.
In our scalar analysis, these are inextricably intertwined.

\noindent{\bf 4.~Renormalons:}\quad
It is thus of great interest to try to extend our
understanding of perturbative quantum field theory by studying the interplay
of coupling-constant and mass renormalization.
In this respect, we noticed an apparent omission,
concerning behaviour at higher orders in the scalar channel.
Let $N_f$ be the number of quark flavours and $\alpha_s(\mu^2)$ be the
strong coupling at scale $\mu^2$. In the limit $N_f\to\infty$,
with $b=N_f\alpha_s/6\pi$ held fixed, the vector correlator
is known~\cite{LNF} to all orders in $b$, at order $1/N_f$.
At first sight, this limit appears
remote from asymptotically free QCD, where the
beta function is dominated by gluons.
However, it has become common practice to
transform large-$N_f$ results to so-called large-$\beta_0$
results, where $\beta_0\equiv(11N_c-2N_f)/12$, with $N_c=3$ colours
and $N_f=3,4,5$ active flavours, gives the one-loop term
in the QCD beta function
\begin{equation}
\beta(\alpha_s)\equiv{d\log\alpha_s\over d\log\mu^2}
=-\sum_{n\ge0}\beta_n\left({\alpha_s\over\pi}\right)^{n+1}\,.
\label{beta}
\end{equation}
By the simplistic device $N_f\to N_f-{33\over2}$, called naive
nonabelianization (NNA) in~\cite{BG}, one transforms
the irrelevant large-$N_f$ ultraviolet (UV) factorial perturbative growth 
of an 
abelian theory,
like QED, into the highly pertinent
large-$\beta_0$ asymptotic perturbative growth  of QCD series, 
since $-b$ is then
replaced by $\overline{b}=\beta_0\alpha_s(\mu^2)/\pi
\approx1/\log(\mu^2/\Lambda^2)$.
Then the so-called renormalon structure (i.e.~the perturbative factorial
growth) of the vector result~\cite{LNF} is related to the way that
long-distance
physics is absorbed into condensates in the operator-product expansion
(OPE) of the vector correlator of QCD~(for the studies in QCD see e.g. 
\cite{Zakh, BY, Chris} and \cite{Alt,BenekeR} for the reviews).

The virtue of an all-orders large-$N_f$ result
is to provide a smooth map from the full two-loop result,
which it exactly reproduces, to the large-order behaviour,
which it reproduces at leading order in $O(1/N_f)$.
It was therefore natural
to inquire whether the vector analysis~\cite{LNF}
had yet been extended to the scalar case. We did not find such a work.
Undertaking the task ourselves, we came to understand why it is
so much more difficult in the scalar channel: the ultraviolet 
(UV) infinities
of mass renormalization must be included to all orders in the coupling.
In Sec.~2, we achieve this, after taking guidance from the
study of critical phenomena.

\noindent{\bf 5.~Multiloop estimates:}\quad
Effort has been expended in estimating effects in electron-positron
annihilation, $\tau$-decay and in deep-inelastic scattering characteristics
beyond the orders of perturbation theory that are exactly computed
(for different approaches see Refs.\cite{KS,SEK,BenekeR}).
A similar attempt was made  
in the case of the decay width of the Higgs boson  \cite{CKS} following the 
ideas of Ref.\cite{KS} and by the authors of Ref.\cite{HPade} 
using the asymptotic  Pad\'e-approximant method of Ref.\cite{EJJKS}.
This can only be inspired guesswork, informed by past experience
and hopeful intuition.
To study in detail the ideas and approximations, lying beyond  such guesswork, 
one must test them 
in detail in as many processes
as possible. In Sec.~3 we submit a variety of suggestions to
detailed scrutiny in the scalar channel, by taking account of existing
4-loop input from~\cite{KC,4LM1,4LM2,4LB}
and by exploiting our new all-orders results at large-$N_f$.
Special attention is paid to the results of application of the 
NNA procedure, closely connected to the large-$N_f$ expansion.
We also investigate the structure of the 3- and 4-loop coefficients 
we are interested in within the ``dual NNA'' procedure, which is 
exact in the large number of colours $N_c$ limit.

\noindent{\bf 6.~Analytical continuation:}\quad
Direct  multiloop calculations are usually performed 
in the euclidean space-like region. However, for the scalar correlator 
the quantities of phenomenological interest, namely the spectral 
functions of the QCD sum rules (see e.g. Refs.\cite{SVZ,ChKrT,BGen})
and the decay width of the Higgs boson are proportional to 
its imaginary part, defined in the minkowskian time-like region. 
In higher orders  the coefficients of perturbative expansions contain 
$\pi^2$-contributions, which generally speaking, are not small 
and can affect the asymptotic structure of the perturbative series 
in powers of $\alpha_s/\pi$. These effects have been much studied 
(see e.g. Refs.\cite{r0}-\cite{r11},\cite{Gor1}) 
in attempts to resum them in all-orders of perturbation theory. 
In Sec.~3, using the results of Sec.~2, we develop further this 
approach to the case of fractional powers of $\alpha_s$, which appear 
in the relation between the $\overline{\rm MS}$-scheme running quark mass 
$m(\mu^2)$ and the scheme-invariant mass $\hat{m}$,  introduced 
in Refs.\cite{r13,Becchi} 
(recent analogous 
independent considerations were given in Ref.\cite{r12}). 
The NNA approximation and the Borel 
resummation technique will be essential theoretical cornerstones of our 
analysis.

In the  Conclusions we discuss the  theoretical uncertainties of the various approaches and 
results, considered in the previous sections, and summarize the 
phenomenological relevance of the results obtained.

\section{Renormalon analysis at large $N_f$}

First, we briefly review the much easier vector case,
where the Ward identity $Z_1=Z_2$ protects the renormalon chain from
UV disturbance, by cancellations between the two-loop
skeletons into which the chain is inserted.
All of these vector methods are necessary here, though they are
not sufficient.
Then we turn to the scalar case, which requires more powerful
techniques, since there is no Ward identity to protect the insertions.
Our aim is to handle both UV-divergent two-loop
skeletons exactly, and to renormalize the mass to all orders
in $N_f\alpha_s$ at large $N_f$,
so as to achieve an
$\overline{\rm MS}$-renormalized result
that connects the full two-loop result to the large-$N_f$
renormalons, checking en route the $O(1/N_f)$ terms in the
3-loop correlator
and 4-loop imaginary part that were given in~\cite{KC}.

In the interests of transparency, we describe the situation
at large-$N_f$, in the first instance, since this is a well-defined
limit. Only later do we allow ourselves the prevailing luxury
of believing that this has anything to do with QCD.
By presenting things this way, we allow the reader to distinguish
hard (and also new) analysis from more easy (and also old)
conjecture.

\subsection{Combinatorics of vector resummation}

Consider the formal series
\begin{equation}
\Pi_V(b,\ep)=-\sum_{n>1}\left({b\over b-\ep}\right)^{n-1}
{L(\ep,n\ep)\over n}
\label{Pbe}
\end{equation}
where the multiloop generating function
\begin{equation}
L(\ep,\de)=\sum_{j,k\ge0}L_{j,k}\ep^j\de^k
\label{Led}
\end{equation}
is regular near $\ep=\de=0$. Here $b$ stands for a renormalized
coupling chosen such that the large-$N_f$ beta function
vanishes at $b=\ep$ in $d\equiv4-2\ep$ dimensions.
In~(\ref{Pbe}), the denominator $b-\ep$
comes from transforming the bare to
the renormalized coupling. For example, in minimally subtracted
large-$N_f$ QED, the bare charge $e_0$ is related to
$b=N_f\alpha(\mu^2)/3\pi$ by
\begin{equation}
{e_0^2\over4\pi^2}={3b\over1-b/\ep}\,
{(\mu^2/4\pi)^\ep\Gamma(1-2\ep)
\over[\Gamma(1-\ep)]^2\Gamma(1+\ep)}
\label{QED}
\end{equation}
where $\mu^2$ makes $b$ dimensionless, and the $\Gamma$ functions
make one-loop massless two-point diagrams rational in $\ep$ at euclidean
momentum $\mu^2=Q^2$.

Now we resum the series by collecting powers of the
renormalized coupling, obtaining
\begin{equation}
\Pi_V(b,\ep)=\sum_{n>1}b^{n-1}\sum_{k<n}{C_{n,k}\over\ep^k}
\label{Pce}
\end{equation}
where the Laurent series at $n>1$ loops starts, by assumption,
with $1/\ep^{n-1}$.
At any order in renormalized perturbation theory,
there are only two significant terms, namely~\cite{LNF}
\begin{eqnarray}
C_{n,1}&=&{L_{n-2,0}\over n(n-1)}\label{C1}\\
C_{n,0}&=&{L_{n-1,0}\over n(n-1)}+(-1)^n(n-2)!L_{0,n-1}\label{C0}
\end{eqnarray}
which are obtained by formal combinatorics.
The first result determines the $n$-loop contribution to the
anomalous dimension; the second gives the finite part.
Thus, in this simple vector case, protected by the Ward
identity $Z_1=Z_2$, it is not necessary to know everything about
the master $n$-loop integral $L(\ep,n\ep)$, resulting from a chain
of $n-2$ fermion loops in a pair of two-loop skeletons. It suffices to know
$L(\ep,0)$ and $L(0,\de)$, i.e.~the analytic continuation to
zero loops, and the Borel limit
$\ep\to0$, with $\de=n\ep$ fixed.

\subsection{Analysis of vector resummation}

In the case of the correlator of the vector current, the analytical problem
was solved in closed form by the first author, who obtained~\cite{LNF}
\begin{eqnarray}
L(\ep,0)&=&{(1+\ep)(1-2\ep)(1-2\ep/3)
\over B(2-\ep,2-\ep)\Gamma(3-\ep)\Gamma(1+\ep)}\,,\label{Le}\\
L(0,\de)&=&\left({\mu^2{\rm e}^{5/3}\over Q^2}\right)^\de
{32\over2-\de}\sum_{k=2}^\infty{(-1)^k k\over (k^2-(1-\de)^2)^2}\,.
\label{Ld}
\end{eqnarray}
The simple $\Gamma$-function result~(\ref{Le})
was long since known from~\cite{pmp}.
The all-orders result~(\ref{Ld}) of~\cite{LNF} is in agreement 
with the independently calculated 
perturbative expansion of Ref.\cite{MB}. At two loops,
\begin{equation}
L(0,0)=16\sum_{k=2}^\infty{(-1)^k k\over(k^2-1)^2}=3
\label{00}
\end{equation}
gives the Jost-Luttinger~\cite{JL} singularity, when $b=\alpha/3\pi$.

Before explaining how to solve the demanding case of the scalar correlator,
we need to explain how the far easier vector case was handled. At $n$-loops,
there are two diagrams: in the first a chain of $n-2$
fermion loops dresses a fermion line; in the second, it is exchanged
between fermion and antifermion. The first case is easy:
only $\Gamma$ functions occur;
the second is hard: there is an $F_{3,2}$ hypergeometric series of the
form given in~\cite{BGK}. However, this series is needed only in the Borel
limit, where it gives a trigamma function.
The series expansion in $\de$ of the 4-dimensional two-point two-loop
scalar diagram, with a modification $(1/k^2)^{1+\de}$ of the
momentum dependence of the totally internal propagator, is~\cite{BGK}
\begin{equation}
I(\de)=8\sum_{l>0}\zeta(2l+1)l(1-4^{-l})\de^{2l-2}
\label{Id}
\end{equation}
with $I(0)=6\zeta(3)$ giving the familiar unmodified result 
(see e.g.\cite{ChT}).
The renormalon series in~(\ref{Ld}) is obtained from the
powerful trigamma identity
\begin{equation}
4(1-\de)\sum_{k=2}^\infty{(-1)^k k\over(k^2-(1-\de)^2)^2}
={1\over(1-\de)^2}-{1\over(2-\de)^2}-{\de\over2}I(\de)
\label{cancel}
\end{equation}
which cancels the singularities of~(\ref{Id}) at both $\de=1$ and $\de=2$.
The vector result~(\ref{Ld}) has double poles
at the negative integers, and at positive
integers greater than 2. Yet it has no pole at $\de=1$ and
merely a single pole at $\de=2$. In the QCD case, the pole
at $\de=2$ signals
the need to absorb long-distance effects into
a nonperturbative gluon condensate, at order $1/Q^4$.
No pole can appear at $\de=1$
since there is no dimension-2 gauge-invariant operator
in the massless theory that could produce nonperturbative effects of
order $1/Q^2$.
The appearance of ${\rm e}^{5/3}$ in~(\ref{Ld}) is easy to understand:
in $d\equiv4-2\ep$ dimensions, each of the $n-2$ one-loop insertions
brings with it a factor
\begin{equation}
f(\ep)\equiv{3\over4}\left({1\over d-1}+{1\over d-3}\right)
=1+{5\ep\over3}+O(\ep^2)
\label{53}
\end{equation}
giving $[f(\ep)]^{n-2}\to\exp(5\de/3)$, in the Borel limit.
To suppress it, one may set $\mu^2=Q^2\exp(-5/3)$,
in the $\overline{\rm MS}$ scheme, corresponding to the perhaps
more physical procedure of subtracting the one-loop photon propagator
at $Q^2$, in the MOM-scheme in QED or $V$-scheme in QCD (for its 2-loop  
definition see Ref.\cite{V}).

\subsection{Combinatoric and analytic complexity in the scalar case}

In the case of the correlator of the scalar current $m\overline{\psi}\psi$
both the combinatorics and the analysis are more demanding.
The behaviour at large $Q^2$ and large $N_f$
is given by
\begin{eqnarray}
\Pi_S&=&\Pi_1(\ep)\left(1+2C_F{I_A(b,\ep)+I_B(b,\ep)\over
T_F N_f}+O(1/Q^2)+O(1/N_f^2)\right)\label{PAB}\\
\Pi_1(\ep)&\equiv&-2[m(\mu^2)]^2Q^2d_F\,{(\mu^2/Q^2)^\ep\over\ep(1-2\ep)}
\label{Pi1}\\
I_A(b,\ep)&\equiv&\int_0^b{g(x)\over\ep-x}\,{\rm d}x\label{A}\\
I_B(b,\ep)&\equiv&\sum_{n>1}\left({b\over b-\ep}\right)^{n-1}
{G(\ep,n\ep)\over n(n-1)}\label{B}
\end{eqnarray}
with $b\equiv T_F N_f \alpha_s(\mu^2)/3\pi$ giving the large-$N_f$
contribution to the beta function. Here $N_f$ is the number of
fermion flavours, $C_F$ is the value of quadratic Casimir operator in the
fundamental fermion representation, $d_F$ is the dimensionality of this
representation, and $T_F$ specifies the normalization of
the corresponding matrices. Concretely, $C_F=4/3$, $d_F=3$,
and $T_F=1/2$, in QCD. In the abelian case of
QED, we simply set $C_F=d_F=T_F=1$.

The stumbling block is the $\overline{\rm MS}$-renormalized mass $m(\mu^2)$
in the one-loop term~(\ref{Pi1}). Renormalization of the scalar vertices
is mandatory: without it the discontinuity of $\Pi_S$ in the physical
(i.e.~minkowski) region $-Q^2=s>0$ would be infinite at $\ep=0$.
At large $N_f$, the multiplicative
renormalization $Z_m=m_0/m(\mu^2)$ may be considered as additive,
with a vertex counterterm $Z_m-1$ giving integral~(\ref{A}).
The numerator $g$ of the integrand is the all-orders contribution to
the anomalous mass dimension
\begin{equation}
\gamma_m(\alpha_s)=
{d\log m(\mu^2)\over d\log(\mu^2)}=
-{C_F b\over T_F N_f}g(b)+O(1/N_f^2)
\label{gm}
\end{equation}
at large $N_f$. The one-loop value
$g(0)=9/4$ gives $\gamma_m(\alpha_s)=-\alpha_s/\pi
+O(\alpha_s^2)$ in QCD.

It is straightforward to insert a chain of fermion loops in the
one-loop diagram for the scalar vertex and obtain the
critical exponent
\begin{equation}
g(\ep)={(d-1)^2\over d}\,
{\Gamma(d-2)\over[\Gamma(d/2)]^2}\,
{\sin\pi\ep\over\pi\ep}
\label{gep}
\end{equation}
at large $N_f$ in $d\equiv4-2\ep$ dimensions,
giving the all-orders result for $g(b)$ in~(\ref{gm}),
by virtue of the fixed point at $b=\ep$, to leading order
in $1/N_f$. We note, for future reference, that $g(\ep)$ is finite
for all $d>-1$, and hence that $g(b)$ is finite for $b<5/2$.
The validity of~(\ref{gep}), for all $d>-1$, is the true origin of the
one-loop result $g(0)=9/4$, at $d=4$.

Anomalous dimensions at $O(1/N_f)$ are given by functions that differ
from~(\ref{gep}) only by multiplication of a rational function of $d$.
The $O(1/N_f^2)$ term in~(\ref{gm}) was found
in~\cite{JG1} in the case of QED, and very recently in~\cite{JG2}
for the yet more demanding nonabelian case of QCD. There one finds
derivatives of $\Gamma$ functions, which still give zeta values.
At $O(1/N_f^3)$, hypergeometric series~\cite{BGK} give
multiple zeta values (MZVs), such as $\zeta(5,3)=\sum_{m>n>0}1/m^5n^3$,
in anomalous dimensions.
This irreducible MZV already occurs at $O(1/N_f)$ in the $\ep$-expansion
of the multiloop generator $G(\ep,n\ep)$ in~(\ref{B}).

Even the handling of the vertex renormalization~(\ref{A}), at large $n$,
is nontrivial.
For the $n$-loop term in the scalar correlator, one needs to expand
the $\Gamma$ functions and rationals of~(\ref{gep}) to
order $\ep^{n-2}$, and then make a further Laurent expansion of
the integrand. After integration, this three-fold series
gets multiplied by $\sum_{k\ge0} 2^{k}\ep^{k-1}$,
from the one-loop term. The fifth and final
series is the most potent: one must multiply by
$(\mu^2/Q^2)^\ep=\sum_{k\ge0} (L\ep)^k/k!$ with the
resulting complicated dependence on $L\equiv\log(\mu^2/Q^2)$ required
to cancel nonlocal terms in the Laurent expansion of
the multiloop diagrams.
Thus the innocent-looking integral~(\ref{A}) generates
a combinatoric plethora of products of rationals,
zeta-values, powers of $1/\ep$, powers of the coupling, and powers of logs.
And thus far we speak only of the analytically
tractable term, free of the MZVs
that occur in true multiloop diagrams.

Next one sees, in $I_B$, the astounding interconnectedness of
perturbative quantum field theory: there is a vast conspiracy,
on a scale that would be ludicrous in human affairs,
between analytically nontrivial $n$-loop integrals in~(\ref{B}), where
$G(\ep,n\ep)$ entails the $F_{3,2}$ hypergeometric series of~\cite{BGK},
and the 5-fold series from the intricate combinatorical processing of
$g(\ep)$ by vertex renormalization.
Everyone knows what this conspiracy must achieve: total cancellation
of logs from the singular terms, as required by the locality of counterterms.
How, one asks, is this conspiracy coordinated? Its success is not in doubt:
field theory cannot fail. Yet nothing, prior
to this work, appeared to indicate the analytical mechanism.

Clearly the analysis of the vector case cannot solve this problem.
Both of its key assumptions are now vitiated. First,
the $n$-loop terms in~(\ref{B}) have a form that does not conform
to~(\ref{Pce}); secondly, Laurent expansion of either~(\ref{A}) or~(\ref{B})
generates a $1/\ep^n$ singularity at $n$ loops,
in defiance of the restriction $k<n$ in~(\ref{Pce}).
It was the Ward identity
$Z_1=Z_2$ of QED, and hence of QCD at large $N_f$, that
protected us from these eventualities in the vector case, by giving
one less factor of $1/\ep$, and by allowing an Ansatz~(\ref{Pce}),
with no singularity at $n=1$.

\subsection{Reconciliation in the scalar case}

How is progress possible in this complex scalar case? By virtue of our
recent finding that
\begin{equation}
G(\ep,\ep)=g(\ep)
\label{simple}
\end{equation}
which expresses the remarkable fact
that analytic continuation to $n=1$, of the hypergeometric
result for inserting chains of $n-2$ one-loop diagrams in a pair
of two-loop skeletons, gives the anomalous dimension,
to all orders in the coupling at large $N_f$. To prove~(\ref{simple}), one must first show that the
irreducibly hypergeometric terms in $G(\ep,\de)$ vanish at $\de=\ep$,
hence cancelling the new singularity at $n=1$ in~(\ref{B}),
which was not encountered in the Ward-protected vector analysis,
based on~(\ref{Pbe}).
Thanks to the systematic hypergeometric methods of~\cite{BGK} this is now
possible. The surviving terms in the analytic
continuation to $n=1$ then give $\Gamma$ functions,
multiplied by a very complicated rational function of $\ep$ and
$\de$. At $\de=\ep$, the match of $G(\ep,\ep)$ to the critical exponent
$g(\ep)$ is perfect.

This enables us to organize both the combinatorics and the analysis,
by writing
\begin{equation}
G(\ep,\de)=
g(\ep){G_2(\ep,\de)\over G_1(\ep,\de)}
\left\{1+(\de-\ep)\left[G_E(\ep)+G_D(\de)+\ep\de G_3(\ep,\de)\right]\right\}
\label{GDE}
\end{equation}
with a prefactor specified by
\begin{eqnarray}
G_1(\ep,\de)&\equiv&
\left({\mu^2\over Q^2}\left(f(\ep)\right)^{1/\ep}\right)^{\ep-\de}
\label{G1}\\
G_2(\ep,\de)&\equiv&{\Gamma(1+\de)\over\Gamma(1+\de-2\ep)}\,
{\Gamma(1-\de+\ep)\over\Gamma(1-\de-\ep)}\,
{\Gamma(1-2\ep)\Gamma(1-\ep)\over\Gamma(1+\ep)}\label{G2}
\end{eqnarray}
and functions of $\ep$ and $\de$ that we eventually obtain,
by hypergeometric analysis, as
\begin{eqnarray}
G_E(\ep)&=&\ep(1-2\ep)\label{Ge}\\
G_D(\de)&=&{2\over1-\de}-{1\over2-\de}+
{8(1-\de)\over3}\sum_{k=2}^\infty{(-1)^k k\over(k^2-(1-\de)^2)^2}
\label{Gd}\\
&=&\sum_{k>0}{k+3\over3}(2-2^{-k})\de^{k-1}
-{8\over3}
\sum_{l>0}\zeta(2l+1)l(1-4^{-l})\de^{2l-1}
\label{Gdex}
\end{eqnarray}
with~(\ref{Gd}) showing the renormalon structure and~(\ref{Gdex}) giving
the Taylor expansion about $\de=0$, thanks to~(\ref{Id},\ref{cancel}).
As in the vector case~\cite{LNF,Chris} no even zeta value
can occur in the expansion of the renormalon contributions, since
$I(\de)=I(-\de)$ is conformally invariant. In the new scalar
result~(\ref{Gdex}) one has the further simplification that odd zeta values
occur only in odd Taylor coefficients.
The vastly more complicated residuum, $G_3$, which we also determined
completely, makes no contribution to either the anomalous dimension
or the finite part, since it is multiplied by a factor that vanishes
at $\de=\ep$, $\de=0$, and $\ep=0$, corresponding to $n=1$, $n=0$, and
$n\to\infty$. After extracting unity, $G_E$, and $G_D$,
one may simply throw away whatever remains in $G_3$.
We did this the hard way, by calculating everything, exactly.

The highly coordinated conspiracy~(\ref{simple}) is signalled by the
leading unity in the braces of~(\ref{GDE}), and the triviality
of $G_1(\ep,\ep)=1$. Moreover, the remarkable combination
of $\Gamma$ functions in~(\ref{G2}) gives
$G_2(\ep,\ep)=G_2(\ep,0)=G_2(0,\de)=1$, which means that we may set
$G_2$ to unity, without the slightest effect on the outcome,
for the same reason that we may set $G_3$ to zero.
Next we note that~(\ref{Ge}) precisely cancels the rational denominator of
the one-loop term~(\ref{Pi1}): this renormalon-free term is
as simple as one could ever hope.
Inspecting the analytic structure
of the nontrivial renormalon contribution~(\ref{Gd}) we see single poles
at $\de=1$ and $\de=2$. The pole at $\de=2$ corresponds to the
appearance of a gluon-condensate contribution of order
$m^2Q^2\langle G_{\mu,\nu}G^{\mu,\nu}\rangle/Q^4$
in $\Pi_S$. This was to be expected, by analogy with the vector case.

The {\em totally new feature} is the pole at $\de=1$,
reflecting the ambiguity of the constant term in the correlator,
which is infinite, though formally proportional to $m^4$,
in perturbation theory,
while current algebra relates it to $\langle m\overline\psi\psi\rangle$.
It is wonderful that an analysis of finite parts of
massless diagrams at large loop numbers can so powerfully remind one of
what one knows from massive diagrams~\cite{DJB} at low loop numbers,
namely that the UV physics of the second subtraction
in the dispersion relation is as profound as that of the first.
It makes no sense to say that at large $Q^2$ we can forget about
the second UV subtraction, because $m^2/Q^2$ is small.
If we make believe that we can,
the $\de=1$ infrared (IR) renormalon of massless diagrams
will remind us of the ultraviolet physics of massive diagrams.
It was, of course, the Ward identity that protected us from
this consideration in the vector case.

Proceeding, we see that the closed forms~(\ref{Ge},\ref{Gd})
lead to renormalized contributions that are speedily found,
to very high orders,
by the combinatorics~(\ref{C1},\ref{C0}) that served in the
vector analysis of~\cite{LNF}, where results were
obtained up to 20 loops, analytically, and up to 100 loops, numerically.
It remains, however, to resum
the leading term in~(\ref{GDE}).
We may set $G_2=1$, and write
$1/G_1=(1-G_1)/G_1+1$, with only the final unity requiring
further attention. That is now feasible, since
the formal transformation
\begin{equation}
\sum_{n>1}\left({b\over b-\ep}\right)^{n-1}{1\over n(n-1)}
=1+{\ep\over b}\log\left(1-{b\over\ep}\right)
=-\sum_{n>1}\left({b\over\ep}\right)^{n-1}{1\over n}
\label{easy}
\end{equation}
gives a rather simple Laurent series, when multiplied by
$g(\ep)=\sum_{n>0}g_n\ep^{n-1}$. Finally, we expand $1/(\ep-x)$
in $x/\ep$ and formally integrate~(\ref{A}).

Tidying up, we see that $I_A+I_B$ is equivalent to $J_A+J_B$,
as far as pole terms and finite parts are concerned, where
\begin{equation}
J_A(b,\ep)\equiv\sum_{n>1}b^{n-1}\left\{
\sum_{j=1}^{n-1}{g_{n-j}\ep^{-j}\over n(n-1)}
-\sum_{k=0}^{\infty}{g_{n+k}\ep^{k}\over n}\right\}
\label{JA}
\end{equation}
combines $I_A$ with the recalcitrant leading term from $I_B$,
and the remainder of $I_B$ is equivalent to
\begin{equation}
J_B(b,\ep)\equiv\ep g(\ep)
\sum_{n>1}\left({b\over b-\ep}\right)^{n-1}
{L_B(\ep,n\ep)\over G_1(\ep,n\ep)}{1\over n}
\label{JB}
\end{equation}
where
\begin{equation}
L_B(\ep,\de)\equiv{G_1(\ep,\de)-1\over\ep-\de}+\ep(1-2\ep)+G_D(\de)
\label{LB}
\end{equation}
is easily
processed by~(\ref{C1},\ref{C0}), using~(\ref{gep},\ref{G1},\ref{Gd}).
It is important to note that $J_A$ is no longer minimal: the first sum
combines pure pole terms of $I_A$, with weight $1/(n-1)$, and the pole
terms from multiplication of $g(\ep)$ by~(\ref{easy}), with weight
$-1/n$. Then a regular part appears, in the second series
of $J_A$. The overall $1/\ep$ singularity, from the one-loop diagram,
means that the $k=0$ term in~(\ref{JA}) becomes singular,
and the $k=1$ term becomes finite.
This will be seen to be crucial, in the following analysis.

\subsection{Critical behaviour of the scalar correlator}

Now we analyze the so-called scalar-scalar anomalous dimension,
$\gamma_{SS}$, defined by~\cite{KC}
\begin{equation}
\left({\partial\over\partial\log\mu^2}+
\beta(\alpha_s){\partial\over\partial\log\alpha_s}
+2\gamma_m(\alpha_s)\right)\Pi_S
=[m(\mu^2)]^2Q^2\left\{\gamma_{SS}(\alpha_s)
+O\left({m^2\over Q^2}\right)\right\}
\label{CZ}
\end{equation}
The presence of an $O(m^2/Q^2)$ term in the braces on the r.h.s.\ of the
renormalization-group equation~(\ref{CZ}) reminds us that {\em two}
subtractions~\cite{Becchi} are required for the scalar correlator. Here
we are concerned with the first, which produces a scale dependence
described by $\gamma_{SS}$. If one supposes that the second might
be forgotten at large $Q^2$, the unity of field theory soon corrects one:
we have already seen that the finite parts of {\em massless}
diagrams at large loop numbers are profoundly aware of the $O(m^2/Q^2)$
UV physics in~(\ref{CZ}), via the $\de=1$ IR renormalon
in~(\ref{Gd}), at $m^2/Q^2=0$.

Working to leading order in $1/Q^2$,
and next-to-leading order in $1/N_f$, we write
\begin{eqnarray}
\Pi_S&=&[m(\mu^2)]^2Q^2d_F
\left(-2L-4+{C_F b\over T_F N_f}
H(L,b)+O(1/Q^2)+O(1/N_f^2)\right)\label{HSS}\\
\gamma_{SS}&=&d_F\left(-2+{C_F b\over T_F N_f}
h(b)+O(1/N_f^2)\right)\label{GSS}
\end{eqnarray}
where $b\equiv T_F N_f\alpha_s/3\pi$ is the leading term
of the 4-dimensional beta function at large $N_f$, i.e.~the value
of $\epsilon$ such that the critical dimension is $4-2\ep$.
Hence the renormalization-group equation~(\ref{CZ})
simplifies to
\begin{equation}
\left({\partial\over\partial L}+
\ep^2{\partial\over\partial\ep}\right)\ep H(L,\ep)
+4(L+2)\ep g(\ep)=\ep h(\ep)
\label{CZe}
\end{equation}
where $L\equiv\log(\mu^2/Q^2)$.
Our aim is to determine $h(\ep)$
to all orders. Then the dependence of $H(L,b)$ on $L$
is completely determined by $H(0,b)$.

We stress the underlying principle
of this work by using the argument $\ep$ in~(\ref{CZe}),
where a reader more used to 4-dimensional perturbation theory
might reasonably expect to see us choose $b\equiv T_F N_f\alpha_s/3\pi$.
The reason should be clear:
the large-$N_f$ beta function is $b-\ep$ in $d\equiv4-2\ep$ dimensions,
where $\ep$ need not be small.
By working near the critical point, $b=\ep$, without assuming that
$d$ is near 4, we bypass perturbation theory.

We have remarked that true anomalous dimensions differ from~(\ref{gep})
only by rational functions of $d$, at order $1/N_f$.
Of course the beta function is not such an object, since it
vanishes at the fixed point, by definition. The reason why
the $O(1/N_f)$ corrections to the 4-dimensional
QED beta function are given by the integral of~(\ref{Le}) is clear:
the physics resides in the
critical exponent that is the derivative of this integral.
In the scalar case, the critical
exponent is the derivative of $\ep h(\ep)+4g(\ep)$.
With some ingenuity, one may obtain it by careful parsing
of~(\ref{JA},\ref{LB}),
which yield
\begin{equation}
{1\over g(\ep)}{{\rm d}\over{\rm d}\ep}\left(
{\ep h(\ep)+4g(\ep)\over4}\right)={(d-3)^2\over3}-2\,.
\label{parabola}
\end{equation}
Here, in short order, is the proof of this fine parabola:
\begin{equation}
{1\over d-3}\left\{2-{(f(\ep)-1)/\ep+\ep(1-2\ep)+2\over f(\ep)}\right\}
={1-f(\ep)-\ep^2\over\ep f(\ep)}={(d-3)^2\over3}-2
\label{oneline}
\end{equation}
with a seemingly troublesome singularity at $d=3$
turning out to give a harmless minimum.
The prefactor $1/(d-3)=1/(1-2\ep)$ comes from
the one-loop result~(\ref{Pi1}). The first term in braces
comes from the first series in $J_A$; the remainder from setting
$\de=0$ in $L_B$. The weight of the series is doubled by the
shift $1/\ep(1-2\ep)=1/\ep+2/(1-2\ep)$;
the fourth term follows from using~(\ref{00}) in~(\ref{Gd}),
which gives $G_D(0)=2$.
By taking the derivative of $\ep h(\ep)+4g(\ep)$,
we remove the nonminimal series in $J_A$, whose
weight, $1/n$, is incommensurate with~(\ref{C1}),
which we use for the $J_B$ terms. The rational function $f(\ep)$
enters via~(\ref{G1}). Taking the exact one-loop vector result
from~(\ref{53}), we obtain parabola~(\ref{parabola}).%~$\Box$

Recall what makes this possible:
the circumstance~(\ref{simple})
that analytic continuation of the hypergeometric series of~\cite{BGK}
to $n=1$, i.e.~to {\em minus} one insertions, reproduces the
large-$N_f$ anomalous mass dimension to all orders in the coupling.
We began with two bad problems, for which the vector analysis gave
no preparation: first we had an extra singularity, vitiating
the combinatorics; secondly we had incommensurate weights
for mass renormalization and those terms that we could handle.
The beauty of~(\ref{simple}) is that it enabled us
to solve both problems at the same time: transferring
the combinatorically recalcitrant term to join the mass renormalization
in $J_A$, we obtained the desired weight as $1/(n-1)-1/n=1/n(n-1)$,
in conformity with~(\ref{C1}). There was a price to pay:
this transfer took a nonminimal term with it. But that was no problem:
we knew from critical phenomena that we must take a derivative,
so as to obtain a physically significant exponent.
In that derivative we must include the precise multiple of $g(\ep)$
that kills the nonminimal terms.

The road from analysis of the Saalschutzian
$F_{3,2}$ series of multiloop~\cite{BGK} diagrams,
to the simple parabola~(\ref{parabola}), was a long one.
Along it there was a narrow bridge: $G(\ep,\ep)=g(\ep)$. We have found
no other route.

\subsection{Analytical results at large $N_f$}

Recalling that $b\equiv T_F N_f\alpha_s/3\pi$, we obtain
\begin{equation}
\gamma_{SS}(\alpha_s)=d_F\left(-2+{C_F\alpha_s\over3\pi}
\sum_{n>1}\left(T_F N_f{\alpha_s\over3\pi}\right)^{n-2}\left\{h_n
+O(1/N_f)\right\}\right)\label{gss}
\end{equation}
where $h(\ep)=\sum_{n>1}h_n\ep^{n-2}$. Working to merely 4 loops,
we immediately find
\begin{equation}
h_2=-{15\over2}\,,\quad
h_3=9\,,\quad
h_4=-18\zeta(3)+{1625\over72}\,,
\label{h4}
\end{equation}
in agreement with~\cite{KC}. The development
\begin{eqnarray*}
h_5&=&-27\zeta(4)+{15\over2}\zeta(3)+{1625\over96}\\
h_6&=&-54\zeta(5)+{27\over2}\zeta(4)+{177\over10}\zeta(3)+{8923\over480}\\
h_7&=&-90\zeta(6)+30\zeta(5)+{53\over2}\zeta(4)
-18[\zeta(3)]^2+{593\over18}\zeta(3)+{1955\over96}\\
h_8&=&-162\zeta(7)+{375\over7}\zeta(6)+{741\over14}\zeta(5)
-54\zeta(4)\zeta(3)+{2621\over56}\zeta(4)\\&&{}
+{75\over7}[\zeta(3)]^2+{715\over24}\zeta(3)+{59693\over2688}
\end{eqnarray*}
was obtained , using

\begin{equation}
{h_{n+1}+4g_{n+1}\over4}={4g_{n-2}-4g_{n-1}-5g_n\over3n}
\label{hn}
\end{equation}
which solves~(\ref{parabola}), with
\begin{equation}
\sum_n g_n\ep^{n-1}=\left[4-\sum_{n>1}\left({3\over2^n}
+{n\over2}\right)\ep^{n-2}\right]
\exp\left(\sum_{l>2}{2^l-3-(-1)^l\over l}\zeta(l)\ep^l\right)
\label{gn}
\end{equation}
obtained from~(\ref{gep}).
By this means analytical results to 20 loops, and numerical results
to 100 loops, are readily obtainable.
Since $g(\ep)$ is finite for $\ep<5/2$, so is $h(\ep)$.
Hence the coefficients decrease rather rapidly, with $h_n=O((2/5)^n)$.
For example, we found that $h_{16}\approx1.867\times10^{-5}$.

This convergence is in marked contrast with the behaviour of the
finite parts.
Writing $H(L,b)=\sum_{n>1}H_n(L)b^{n-2}$,
we used~(\ref{JA},\ref{LB}) to derive the all-orders solution
\begin{eqnarray}
n(n-1)H_n(L)&=&n(h_{n+1}-4(L+2)g_n)+4g_{n+1}-9(-1)^n D_n(L)
\label{Hn}\\
\sum_n D_n(L)\de^n/n!&=&\left\{1+\de G_D(\de)\right\}
\exp((L+5/3)\de)
\label{Dn}
\end{eqnarray}
of the renormalization-group equation~(\ref{CZe}).
Two good checks are provided by the vanishing of the
r.h.s.\ of~(\ref{Hn}) at $n=0$ and $n=1$, using
$g_1=9/4$, $h_2=4g_2=-15/2$ and $G_D(0)=2$.
Two stronger checks are provided by known results at two~\cite{DJB}
and three~\cite{KC}  loops.
We obtain from~(\ref{Hn})
the entirety of the former and the large-$N_f$ terms of the latter,
for all $\mu^2/Q^2$. At $\mu^2=Q^2$, we write $H_n\equiv H_n(0)$.
In~\cite{KC}, we find
\begin{equation}
H_2=3\left[-{131\over8}+6\zeta(3)\right]\,,\quad
H_3=9\left[{511\over36}-4\zeta(3)\right]\,,
\label{H3}
\end{equation}
at $\mu^2=Q^2$.

In order to extend explicit results to high loop numbers, and to
analyze their UV and IR renormalon content, we separate
$G_D(\de)=G_-(\de)+G_+(\de)$ into
\begin{eqnarray}
G_-(\de)&=&-{2\over3}\sum_{k>0}{(-1)^k\over(k+\de)^2}\label{Gm}\\
G_+(\de)&=&{2\over1-\de}-{1\over2-\de}+{2\over3}
\sum_{k>2}{(-1)^k\over(k-\de)^2}\label{Gp}
\end{eqnarray}
and expand in $\de$. (Note however that the Taylor expansion of the
{\em total} renormalon contribution is already given with great economy
by~(\ref{Gdex}), with odd zeta values in odd Taylor coefficients.)
Computation of analytical
results to 20 loops takes seconds, for any value of
$L\equiv\log(\mu^2/Q^2)$.
The analytical development to merely 8 loops is given by
\begin{eqnarray*}
H_4&=&
90\zeta(5)
-{27\over4}\zeta(4)
+{157\over4}\zeta(3)
-{499069\over1728}
\\
H_5&=&
-{2304\over5}\zeta(5)
-{45\over4}\zeta(4)
-{4337\over60}\zeta(3)
+{1976311\over1728}
\\
H_6&=&
1701\zeta(7)
-15\zeta(6)
+{14829\over10}\zeta(5)
+{537\over40}\zeta(4)
-3[\zeta(3)]^2
+{54643\over360}\zeta(3)
-{840309103\over155520}
\\
H_7&=&
-{99387\over7}\zeta(7)
-{160\over7}\zeta(6)
-{87026\over21}\zeta(5)
-{54\over7}\zeta(4)\zeta(3)
+{332\over21}\zeta(4)
\\&&{}
-{32\over7}[\zeta(3)]^2
-{41614\over189}\zeta(3)
+{1275753995\over40824}
\\
H_8&=&
68850\zeta(9)
-{1323\over32}\zeta(8)
+{3967083\over56}\zeta(7)
+{3635\over112}\zeta(6)
-{27\over2}\zeta(5)\zeta(3)
+{7017943\over672}\zeta(5)
\\&&{}
-{639\over56}\zeta(4)\zeta(3)
+{12973\over896}\zeta(4)
+{727\over112}[\zeta(3)]^2
+{28819423\over72576}\zeta(3)
-{315995418895\over1492992}
\end{eqnarray*}

\subsection{The asymptotics of naive nonabelianization}

We reiterate the often rehearsed (yet never properly justified)
argument for Borel resummability
of large-$N_f$ singularities at $\de<0$, in the case of QCD.
At truly large $N_f$, they give sign-constant series.
Now we imagine that, in some
vague\footnote{It {\em must} be vague. Unlike quark loops,
gluon loops are not gauge invariant.} sense,
the gluon loops ``follow'' the quark loops. Hence we perform the
naive nonabelianization~\cite{BG} $N_f\to N_f-{11N_c\over2}$ ($N_c=3$), which
gives $b\to-\beta_0\alpha_s(\mu^2)/\pi$,
with $\beta_0=(11N_c-2N_f)/12$
giving the relative contributions of gluons and quarks in the one-loop
beta function of QCD. In the real word, with $N_f\le6$ active quarks,
we have $\beta_0>0$ and $\overline{b}\equiv\beta_0\alpha_s(\mu)/\pi\approx
1/\log(Q^2/\Lambda^2)$, if we suppress large logarithms by renormalizing at
$\mu^2=Q^2$. By this sleight of hand, the singularities
at $\de<0$ now give a sign-alternating asymptotic series
that is resummable by Laplace transformation.
A singularity at $\de>0$ leads to an intrinsic ambiguity of order
$\exp(-\de/\overline{b})=O(Q^{-2\de})$
in a dimensionless correlator. In the vector case~(\ref{Gd})
the dominant ambiguity, at $\de=2$, is associated with the gluon condensate.
Here, in the scalar case with $\Pi_S=O(Q^2)$,
a singularity at $\de>0$ leads
to an ambiguity of order $Q^{2-2\de}$.
Hence the renormalon at $\de=1$ is in accord with the
fact that the constant term in $\Pi_S$ is inaccessible to
perturbation theory. Interestingly, our analysis does not
distinguish the scalar from the pseudoscalar channel.
In the latter case, current algebra relates the constant
term in the correlator to the quark condensate
$\langle m\overline\psi\psi\rangle$.

At any given order $n>1$, we may separate $H_n$ into 6 parts:
\begin{equation}
H_n=H^{\overline{\rm MS}}_n+H_n^{(0)}+H_n^{(-)}
+H_n^{(1)}+H_n^{(2)}+H_n^{(+)}
\label{Hr}
\end{equation}
where
\begin{equation}
H^{\overline{\rm MS}}_n={n(h_{n+1}-4(L+2)g_n)+4g_{n+1}\over n(n-1)}
\label{Hbar}
\end{equation}
is the highly convergent part from $\overline{\rm MS}$ renormalization
and $H_n^{(0)}$
comes from the leading unity in the braces of~(\ref{Dn}).
The next term comes from~(\ref{Gm}), with resummable
UV renormalon 
singularities at $\de<0$;
final three terms come from the IR renormalons in~(\ref{Gp}).

Table~1 shows the numerics of this breakdown, again at $\mu^2=Q^2$.
We comment on each contribution in turn.
\begin{enumerate}
\item
The $\overline{\rm MS}$-specific contribution
$H^{\overline{\rm MS}}_n=O((2/5)^n)$
is negligible at large $n$. This is because
the critical exponent~(\ref{gep}) is finite at all dimensions $d>-1$.
The numerator of~(\ref{Le}) shows that the same applies
to the critical exponent in the vector case.
\item
The modest growth of $H_n^{(0)}$
comes from the choice $\mu^2=Q^2$, made so as to compare~(\ref{H3})
with~\cite{KC}.
At $\mu^2=Q^2\exp(-5/3)$, corresponding to QED MOM- or V-schemes
subtraction of quark-loop insertions, these terms fall off like $1/n^2$.
\item
The series coming from $H_n^{(-)}$
is now regarded as infrared-safe: one may resum it by Laplace
transformation of the Borel transform~(\ref{Gm}), obtaining
\begin{equation}
\sum_{n>1}H_n^{(-)}(L)[-\overline{b}]^{n-1}
=9\int_0^\infty{G_-(\de)\exp((5/3+L)\de)-G_-(0)\over\de}
\exp(-\de/\overline{b})\,{\rm d}\de
\label{Hm}
\end{equation}
where $\overline{b}\approx1/\log(\mu^2/\Lambda^2)$ replaces $-b$,
by naive nonabelianization.
\item
The total is dominated by the factorial growth of
$H_n^{(1)}=O((n-2)!)$. Indeed this dominant $\de=1$
IR renormalon even gives a reasonable
account at very low orders. 
For example,
it gives a fraction $73.0/84.5=86\%$
of the 3-loop large-$N_f$ result of~\cite{KC}.
Clearly perturbation theory would be in bad shape, if this
renormalon entered phenomenology.
Fortunately it does not. Rather it is a reminder, from the perturbative
sector, of the long-distance physics in $\Pi_S(0)$, which current
algebra relates to quark condensates. In the sum-rule analysis
of~\cite{CPS}, it was nullified by taking a twice-subtracted
dispersion relation. In Higgs decay and spectral function 
of the QCD sum rules, it is nullified by taking
the imaginary part, in the physical region $s=-Q^2>0$.

As previously remarked, one
must set aside the temptation to divide $\Pi_S$ by $Q^2$ and then
differentiate w.r.t.~$Q^2$,
so as to remove $\gamma_{SS}$ from~(\ref{CZ}). That would {\em not}
remove the $\de=1$ renormalon, since it would leave the kinematic
singularity $\Pi_S(0)/Q^4$. Then long-distance physics would make
the asymptotic perturbation series expire sooner than
needs be, since it is always
faithful to the motto {\sl dulce et decorum est pro Wilson mori}:
it's OK to  
explode in accordance with Wilson's OPE.
Indeed, it often signals impending doom at the 2-loop level~\cite{BG}.
\item
More long-distance physics resides in $H_n^{(2)}=O((n-2)!/2^n)$,
which signals the presence of the gluon condensate in Wilson's
scheme of things, as articulated in QCD by~\cite{SVZ}.
This $\de=2$ IR renormalon is also absent
from the imaginary part, to leading order in $1/s$.
When taking an imaginary part, in the physical
region, one kills any single-pole renormalon, and turns a double-pole
into a single pole, since the imaginary part of~(\ref{Dn})
acquires a factor $\sin(\pi\de)$.
The quark and gluon dimension-4
operators appear in the energy momentum tensor and are hence
renormalization-group invariants, to leading order. Thus they
are absent from the imaginary part, at order $1/\beta_0$,
in the high-energy limit. Since there is
no matrix element to absorb renormalons
at $\de=1$ or $\de=2$, these, and only these,
appear as single poles in~(\ref{Gd}).
\item
The only IR (i.e.~unresummable)
large-$\beta_0$ renormalons that appear in the imaginary part
at large $s$, are those in $H_n^{(+)}$, at $\de>2$.
These correspond to long-distance physics in matrix elements of
operators $O_k$ with dimensions $d_k\ge6$ in the OPE.
The resultant ambiguity in ${\rm Im}~\Pi_S$
is of order $m^2\langle O_k\rangle/s^{d_k-1}$.
\end{enumerate}

\subsection{Analysis of the imaginary part at large $N_f$}

Table~1 might appear alarming. A far happier picture emerges in Table~2,
where we analyze a {\em physical} quantity, namely the imaginary part
${\rm Im}~\Pi_S=2\pi s R_S(1+O(1/s))$
at $-Q^2=s\equiv w^2$. For $w=M_H$, this contains the radiative corrections
to decay of a Higgs boson of mass $M_H$ into a quark-antiquark pair,
ignoring terms of order $(m(M_H^2)/M_H)^2$. Now we are dealing with
a multiplicatively renormalized quantity: the explicit
dependence of $R_S$ on $\mu^2$
is cancelled by its implicit dependence, via $\alpha_s(\mu^2)$
and $m(\mu^2)$, giving
\begin{equation}
\left({\partial\over\partial\log\mu^2}
+\beta(\alpha_s){\partial\over\partial\log\alpha_s}
+2\gamma_m(\alpha_s)\right)R_S=0
\label{CSR}
\end{equation}
with $2\gamma_m$ appearing because $R_S$ contains two powers of the
renormalized mass in the Born approximation. 
We should, however, set $\mu^2=O(s)$,
to suppress large logarithms. 
Here, we set $\mu^2=s\equiv w^2$ and obtain
\begin{equation}
R_S=3[m(w)]^2\left(1+{\alpha_s\over\pi}\sum_{n>1}b_S^{n-2}\left\{
S_n+O(1/N_f)\right\}\right)
\label{RH}
\end{equation}
with $b_S\equiv-N_f\alpha_s(w^2)/6\pi$, which is replaced by
$\beta_0\alpha_s(w^2)/\pi\approx1/\log(w^2/\Lambda^2)$ in naive
nonabelianization.
We have $S_2=17/3$, with the large-$N_f$ term giving the total
radiative correction at 2 loops.
At $n\ge2$ loops, we obtain
the leading term at large-$N_f$ as
\begin{eqnarray}
S_n&=&A_n+\Delta_n\label{Sn}\\
A_n&=&{2(-1)^n\over n-1}\,{g_n\over g_1}\label{An}\\
\sum_{n\ge2}{\de^{n-1}\Delta_n\over(n-2)!}&=&2\exp(5\de/3)
\left\{1+\de G_-(\de)+\de G_+(\de)\right\}{\sin(\pi\de)\over\pi\de}-2
\label{Dpn}
\end{eqnarray}
where $g_n$ is the $n$-loop term in the large-$N_f$ result~(\ref{gn})
for the anomalous quark-mass dimension. The separation $S_n=A_n+\Delta_n$
into anomalous-dimension and renormalon contributions will be used
in Sec.~3, where we shall retain only the latter in large-$\beta_0$
approximations. Here, we retain both terms, so as to present the
exact analytical results at large-$N_f$.

We note that the high-energy imaginary part $R_S$
receives powers of $\pi^2$ from two
sources: from the even zeta values $\{\zeta(2k)\mid k>1\}$ of euclidean
analysis, and also from the analytic continuation of logarithms
to the physical region. At large-$N_f$, the separation of these two
effects is particularly clean: even zeta values occur only in the
anomalous-dimension contributions $A_n$; powers of $\pi^2$ from analytic
continuation result only from the factor $\sin(\pi\de)/\pi\de$
in the Borel transform~(\ref{Dpn}) of the renormalon contributions $\Delta_n$.
We write the $A_n$ terms in braces in the
following large-$N_f$ results. Up to 4 loops we obtain
\begin{eqnarray}
S_2&=&\left\{-{5\over3}\right\}+{22\over3}={17\over3}\label{S2}\\
S_3&=&\left\{{35\over36}\right\}
-4\zeta(3)
-{1\over3}\pi^2
+{275\over18}=
-4\zeta(3)
-{1\over3}\pi^2
+{65\over4}\label{S3}\\
S_4&=&\left\{{4\over3}\zeta(3)-{83\over108}\right\}
-{40\over3}\zeta(3)
-{22\over9}\pi^2
+{3940\over81}=
-12\zeta(3)
-{22\over9}\pi^2
+{15511\over324}\label{S4}
\end{eqnarray}
with totals in agreement the with the large-$N_f$ 3-loop \cite{Gor1}
and 4-loop~\cite{KC} terms.
At 5 and 6 loops our new results
\begin{eqnarray}
S_5&=&A_5+\Delta_5=
\left\{-{3\over2}\zeta(4)+{5\over6}\zeta(3)+{65\over96}\right\}
\nonumber\\&&{}
-60\zeta(5)
+4\zeta(3)\pi^2
-{100\over3}\zeta(3)
+{1\over10}\pi^4
-{275\over18}\pi^2
+{64877\over324}
\label{S5}\\
S_6&=&A_6+\Delta_6=\left\{{12\over15}\zeta(5)-\zeta(4)-{7\over9}\zeta(3)
-{451\over720}\right\}
\nonumber\\&&{}
-400\zeta(5)
+{80\over3}\zeta(3)\pi^2
-{2000\over27}\zeta(3)
+{22\over15}\pi^4
-{7880\over81}\pi^2
+{244871\over243}
\label{S6}
\end{eqnarray}
were found by expanding~(\ref{gn}), to obtain $A_n$ in the braces,
and the new all-orders renormalon results~(\ref{Gm},\ref{Gp}),
to obtain $\Delta_n$ via~(\ref{Dpn}). This method may easily be
continued up to 20 loops, analytically, and up to 100 loops, numerically.

The corresponding vector quantities in electron-positron annihilation,
at large $N_f$, come from the old result~(\ref{Ld}) of~\cite{LNF},
which immediately gives
\begin{equation}
\sum_{n>1}{\de^{n-2}V_n\over(n-2)!}
={8\exp(5\de/3)\over3(1-\de)(2-\de)}
\left(-\sum_{k>0}{(-1)^k\over(k+\de)^2}
+\sum_{k>2}{(-1)^k\over(k-\de)^2}\right)
{\sin(\pi\de)\over\pi\de}
\label{V_n}
\end{equation}
At 6 loops the vector coefficient has grown by an order of magnitude and
changed sign, with $V_6/V_2\approx-11$.
By contrast, $S_6/S_2\approx0.5$ in the scalar channel. This remarkable
postponement of factorial growth, at large-$N_f$, does
not depend on a cancellation between anomalous-dimension and renormalon
terms; rather it reflects cancellations between the renormalons
themselves, with
$\Delta_6/\Delta_2\approx0.4$ showing no sign of the
growth that had become clear at 6 loops in the vector channel.
In this respect, the (pseudo-)scalar channel is better behaved than the
(axial-)vector channel, at high energy and large $N_f$,
despite warnings~\cite{NSVZ} that might suggest an opposite
situation.

\subsection{Postponed factorial growth}

Considerable interest attaches to the numerics of Table~2,
where it will be seen than $S_n$ is amazingly well-behaved for $n<7$,
while $H_n$ in Table~1 had already gone haywire at $n=3$.
We have discovered a plateau of tranquility at loop numbers $n=2,3,4,5,6$,
in the large-$N_f$ terms of the imaginary part
of the scalar correlator.

The behaviour at $n>7$ in Table~2 is fairly clear.
\begin{enumerate}
\item The $\overline{\rm MS}$ term $S_n^{\overline{\rm MS}}$ 
falls of like $(2/5)^n$.
\item The leading unity of the braces of~(\ref{Dpn}) gives 
power growth of $S_n^{(0)}$,
modulated by a sine.
\item Eventually, the UV renormalons at $\de<0$ take over,
giving an alternating series for $S_n^{(-)}$, 
now that we have naively nonabelianized.
But they takes ages to get going: even at $n=8$ they have not yet overtaken
the humble $\de=0$ term.
\item The term $S_n^{(1)}$ from $\de=1$ is no longer a renormalon; 
the single pole has
been cancelled by $\sin(\pi\de)$. 
\item The same applies to the $\de=2$ term $S_n^{(2)}$, except that it is, 
on average,
smaller than the $\de=1$ term, by a factor of order $1/2^n$, after allowing
for the sinusoidal oscillations.
\item The unresummable series with the 
coefficients $S_n^{(+)}$ from the IR renormalons at $\de\ge3$
grows factorially, eventually. However it is suppressed by a factor
of order $1/3^n$, in comparison with the UV renormalons.
\end{enumerate}

The staggering feature is the tally, $S_n$.
For $n>7$, it behaves like the wild animal that it truly is;
for $n<7$, all is sweetness and light.
Presented with only the analytical expressions for $S_n$ at
$n=2,3,4,5,6$, one would not have the slightest inkling of what is
in store. Conversely, one may say that the
large-$N_f$ renormalons show themselves
mercifully late, in this physical quantity.

 Note that these statements are entirely dependent on the scheme
chosen (for a  discussion  of the scheme-dependence 
of renormalon contributions see e.g. Refs.\cite{KrPiv3,BenekeR}).
As will be clear from the discussions of  Sec. 3.5 and Table 5, 
the $V$-scheme results
are in fact  a much better indicator of the eventual asymptotics of the
coefficients. Nonetheless given the widespread use of the
$\overline{\rm MS}$ scheme in the literature we initially focus here
on the asymptotic behaviour in that scheme.

Now consider the case where $s=w^2$ with $w\approx2$~GeV,
as in strange-quark mass extraction~\cite{CPS,CDPS}.
With $\beta_0=9/4$, the expansion
parameter $\beta_0\alpha_s(w)/\pi\approx1/\log(w^2/\mu^2)\approx0.2$
is now uncomfortably large.
Hence the tranquil plateau at $n<7$ loops is good news
for the later analysis in~\cite{CPS}, where
4-loop perturbative QCD was used, on both sides of the sum rule.
The result was significantly different from an earlier
3-loop analysis in~\cite{CDPS}, by the same group,
using a different truncation procedure (compare with the similar 
3-loop studies of Ref.\cite{JM}).

The villains, which might have been waiting just round the corner,
were the renormalons. Might a large 5-loop term significantly
change the 4-loop result?

To date, we know of only one analytical technique for estimating
such effects on the basis of genuinely new calculation,
instead of reshuffling old input: naive nonabelianization
of the large-$N_f$ terms, which we have computed
in the demanding scalar channel, at some cost of labour.
The good news is that we found nothing alarming at $n=5$ loops.
The even better news is that all seems well at $n=6$ loops.
Only at $n=7$ does the inevitable growth show signs of commencing.
Hence the best indicator that we can compute suggests that the
perturbative part of the strange-quark-mass extraction in~\cite{CPS}
is in fine shape.

Indeed, the results of Table 2 indicate that the contribution 
of the 5-loop coefficient 
might be not crucial and that the corresponding 
perturbative series should be truncated at the 6-loop level 
in accordance with the common practice  in treating the
predictions of asymptotic perturbative 
expansions, which presumes their truncation at the minimal term.

This fits nicely with the claim in~\cite{CPS} that
the condensate contributions are also under control. Had these
contributions been substantial, our discovery of perturbative
tranquility at large $N_f$ would have been rather puzzling; now
it may be taken as gratifying evidence of the depth to which the OPE
connects ultraviolet and infrared physics.

\subsection{Euclidean analysis at large $N_f$}

In the vector channel we may express the results
of high-energy perturbation theory in two ways:
either in terms of $R(s)$, in the
physical region, or in terms the Adler function
\begin{equation}
D(Q^2)\equiv Q^2\int_0^\infty{R(s)ds\over(s+Q^2)^2}
\label{D}
\end{equation}
in the euclidean region. Here, $R(s)=1+\alpha_s/\pi+O(\alpha_s^2)$
gives the high-energy
radiative corrections to the parton model in electron-positron
annihilation and hence $D(Q^2)$ gives the corresponding
radiative corrections
to the derivative of the polarization function of the vector correlator.

In both cases, we ignore
quark masses, so the transformation between one set of radiative
corrections and the other is generated, to all orders,
by the following integral
\begin{equation}
Q^2\int_0^\infty{ds\over(s+Q^2)^2}\left({\mu^2\over s}\right)^\de
={\pi\de\over\sin(\pi\de)}
\left({\mu^2\over Q^2}\right)^\de
\label{pi2}
\end{equation}
with the expansion of
\begin{equation}
{\pi\de\over\sin(\pi\de)}=1+\sum_{k>0}\left(2-4^{1-k}\right)
\zeta(2k)\de^{2k}
\label{zeta}
\end{equation}
telling one precisely how to remove from the imaginary part
all and only those powers of $\pi^2$ that
came from analytic continuation of logarithms.

In the scalar channel, we are confronted by a choice of euclideanizations
of the high-energy imaginary part.
The most prudent choice would appear to be the dispersion relation
for the {\em second} derivative of the scalar correlator
\cite{Becchi}, which
is multiplicatively renormalized for all values of $m^2/Q^2$
and is hence free of the IR renormalon
at $\de=1$ in~(\ref{Gp}). Ignoring terms of order $m^2/Q^2$, this amounts
to the euclideanization
\begin{equation}
\overline{D}_S(Q^2)=2Q^2\int_0^\infty
{sR_S(s)ds\over(s+Q^2)^3}
=3[m(Q^2)]^2\left(1+{11\over3}{\alpha_s(Q^2)\over\pi}+O(\alpha_s^2)\right)
\label{Db}
\end{equation}
of the radiative corrections to the imaginary part at high energy.
However, a merely mathematical analogy with the vector case might lead one
to consider the construct
\begin{equation}
\widetilde{D}_S(Q^2)=Q^2\int_0^\infty
{R_S(s)ds\over(s+Q^2)^2}
=3[m(Q^2)]^2\left(1+{17\over3}{\alpha_s(Q^2)\over\pi}+O(\alpha_s^2)\right)
\label{Dt}
\end{equation}
corresponding to a dispersion relation for the first derivative of
$\Pi_S(Q^2)/Q^2$. Here one expects the asymptotic perturbation
series to destroy itself earlier,
leaving an ambiguity of order $\Lambda^2/Q^2$ that reflects the
failure to remove the infinities in $\Pi_S(0)$. 
Therefore, this ambiguity has a perturbative origin.

We shall show that
at large-$N_f$ the perturbative series for 
the nonstandard euclideanization $\widetilde{D}_S$
is indeed worse behaved than the twice-differentiated euclideanization
$\overline{D}_S$, in accord with the expectation from the OPE.
It might therefore be expected that we shall proceed in Sec.~3 only
with the safer alternative $\overline{D}_S$, as used in
QCD sum-rules~\cite{Becchi,CPS}. In fact, we shall need both
constructs, so as to study the logarithmic derivative 
of $\widetilde{D}_S$ , which is the euclidean analog of the considered 
in 
the case of Higgs decay quantity (see  Refs.\cite{Gor3,4LM2,GK}). 
In our case it can be defined as 
\begin{equation}
R_D(Q^2)\equiv-{1\over2}{d\log\widetilde{D}_S(Q^2)\over d\log Q^2}
={\widetilde{D}_S(Q^2)-\overline{D}_S(Q^2)\over2\widetilde{D}_S(Q^2)}
={\alpha_s(Q^2)\over\pi}+O(\alpha_s^2)
\label{RD}
\end{equation}
which satisfies a renormalization-group equation
\begin{equation}
\left({\partial\over\partial\log\mu^2}
+\beta(\alpha_s){\partial\over\partial\log\alpha_s}\right)R_D=0
\label{RGD}
\end{equation}
that is free of the anomalous quark-mass dimension and hence
suitable for analysis by the method of effective charges~\cite{Grunberg1} 
(see also Ref.\cite{Kr}),
scheme-invariant perturbation theory \cite{SIPT}, commensurate 
scale relations \cite{BL} and the 
standard PMS approach \cite{PMS}.

Here we assemble everything that is known about the $\overline{\rm MS}$
perturbation series of the euclidean constructs~(\ref{Db},\ref{Dt},\ref{RD}).
It is convenient to begin with
\begin{eqnarray}
\widetilde{D}_S(Q^2)&=&3[m(Q^2)]^2\left(1+\sum_{n>0} d_n
\left({\alpha_s(Q^2)\over\pi}\right)^n\right)\label{dn}\\
d_1&=&\frac{17}{3}\label{d1}\\
d_2&=&\frac{10801}{144}-\frac{39}{2}\zeta(3)
-\bigg(\frac{65}{24}-\frac{2}{3}\zeta(3)\bigg)N_f
\label{d2}\\
d_3&=&\frac{6163613}{5184}-\frac{109735}{216}\zeta(3)
+\frac{815}{12}\zeta(5)\nonumber\\&&{}
-\bigg(\frac{46147}{486}-\frac{262}{9}\zeta(3)+\frac{5}{6}\zeta(4)
+\frac{25}{9}\zeta(5)\bigg)N_f
+\bigg(\frac{15511}{11664}-\frac{1}{3}\zeta(3)\bigg)N_f^2
\quad{}\label{d3}
\end{eqnarray}
obtained by removing terms involving $\pi^2$
in the expansion
\begin{equation}
R_S(w^2)=3[m(w^2)]^2
\left(1+\sum_{n>0} s_n \left({\alpha_s(w)\over\pi}\right)^n\right)
\label{Rsn}
\end{equation}
of the imaginary part, given to 4 loops in~\cite{KC}.
To effect the inverse transformation, to 5 loops, one may use 
the fixed-order  perturbative expansion in the minkowskian region 
\begin{eqnarray}
s_1&=&d_1\label{s1}\\
s_2&=&d_2-\gamma_0(\beta_0+2\gamma_0)\pi^2/3
\label{s2}\\
s_3&=&d_3-\big[d_1(\beta_0+\gamma_0)(\beta_0+2\gamma_0)
+\beta_1\gamma_0+2\gamma_1(\beta_0+2\gamma_0)\big]\pi^2/3
\label{s3}\\
s_4&=&d_4-\big[d_2(\beta_0+\gamma_0)(3\beta_0+2\gamma_0)
+d_1\beta_1(5\beta_0+6\gamma_0)/2
+4d_1\gamma_1(\beta_0+\gamma_0)
\nonumber\\&&{}
+\beta_2\gamma_0+2\gamma_1(\beta_1+\gamma_1)
+\gamma_2(3\beta_0+4\gamma_0)\big]\pi^2/3\nonumber\\&&{}
+\gamma_0(\beta_0+\gamma_0)
(\beta_0+2\gamma_0)(3\beta_0+2\gamma_0)\pi^4/30
\label{s4}
\end{eqnarray}
where the relation between $s_4$ and $d_4$ was derived in Ref.\cite{CKS} and
\begin{equation}
\gamma_m(\alpha_s)\equiv{d\log m\over d\log\mu^2}
=-\sum_{n\ge0}\gamma_n\left({\alpha_s\over\pi}\right)^{n+1}
\label{gamma}
\end{equation}
gives the expansion of the anomalous quark-mass dimension,
in the same manner that~(\ref{beta}) gives the expansion of the
beta function for the scale dependence of the coupling. Both expansions
are known to 4 loops. The coefficients of $\gamma_m$
are \cite{4LM1,4LM2}:
\begin{eqnarray}
\gamma_0&=&1\label{g0}\\
\gamma_1&=&\frac{1}{16}\bigg[\frac{202}{3}-\frac{20}{9}N_f\bigg]
\label{g1}\\
\gamma_2&=&\frac{1}{64}\bigg[1249-\bigg(\frac{2216}{27}
+\frac{160}{3}\zeta(3)\bigg)N_f-\frac{140}{81}N_f^2\bigg]
\label{g2}\\
\gamma_3&=&\frac{1}{256}\bigg[
\frac{4603055}{162}+\frac{135680}{27}\zeta(3)-8800\zeta(5)\nonumber\\&&{}
-\bigg(\frac{91723}{27}+\frac{34192}{9}\zeta(3)-880\zeta(4)
-\frac{18400}{9}\zeta(5)\bigg)N_f\nonumber\\&&{}
+\bigg(\frac{5242}{243}+\frac{800}{9}\zeta(5)-\frac{160}{3}\zeta(4)\bigg)
N_f^2-\bigg(\frac{332}{243}-\frac{64}{27}\zeta(3)\bigg)N_f^3\bigg]
\label{g3}
\end{eqnarray}
while those of beta function are \cite{4LB}:
\begin{eqnarray}
\beta_0&=&\frac{1}{4}\bigg[11-\frac{2}{3}N_f\bigg]\label{b0}\\
\beta_1&=&\frac{1}{16}\bigg[102-\frac{38}{3}N_f\bigg]
\label{b1}\\
\beta_2&=&\frac{1}{64}
\bigg[\frac{2857}{2}-\frac{5033}{18}N_f+\frac{325}{54}N_f^2\bigg]
\label{b2}\\
\beta_3&=&\frac{1}{256}\bigg[\frac{149753}{6}+3564\zeta(3)
-\bigg(\frac{1078361}{162}+\frac{6508}{27}\zeta(3)\bigg)N_f\nonumber\\&&{}
+\bigg(\frac{50065}{162}+\frac{6472}{81}\zeta(3)\bigg)N_f^2
+\frac{1093}{729}N_f^3\bigg]
\label{b3}
\end{eqnarray}
which will likewise be needed in our analysis.

The results at large-$N_f$ are
\begin{eqnarray}
d_{n-1}&=&(-N_f/6)^{n-2}\left(A_n+\widetilde{\Delta}_n+O(1/N_f)\right)
\label{dtl}\\
\sum_{n\ge2}{\delta^{n-1}\widetilde{\Delta}_n\over(n-2)!}&=&2\exp(5\de/3)
\left\{1+\de G_-(\de)+\de G_+(\de)\right\}-2
\label{Dtn}
\end{eqnarray}
with the Borel transform~(\ref{Dtn}) giving
\begin{eqnarray}
\widetilde{\Delta}_5&=&
-60\zeta(5)
-{100\over3}\zeta(3)
+{64877\over324}
\label{Dt5}\\
\widetilde{\Delta}_6&=&
-400\zeta(5)
-{2000\over27}\zeta(3)
+{244871\over243}
\label{Dt6}
\end{eqnarray}
at 5 and 6 loops.

We expect the factorial growth of~(\ref{dtl}),
in $\widetilde{D}_S$, to be more drastic
than that of
\begin{equation}
s_{n-1}=(-N_f/6)^{n-2}\left(A_n+\Delta_n+O(1/N_f)\right)
\label{sl}
\end{equation}
in the imaginary part $R_S$,
since the latter is free of the spurious $\de=1$ renormalon
that afflicts the former. Table~3 emphatically confirms this
expectation at large $N_f$, where one sees that
$\widetilde{\Delta}_6/\widetilde{\Delta}_2\approx69$ is two orders of
magnitude larger than $\Delta_6/\Delta_2\approx0.40$.

In general, one expects that the asymptotic structure  of
perturbation theory expansions will differ for the 
physical quantity $R_S(s)$ and the euclideanization
$\widetilde{D}_S(Q^2)$.
To construct the latter from the former one takes the renormalization-group
determined powers of the minkowski logarithm $\log(\mu^2/w^2)$ and performs
the transformation
\begin{equation}
\log^{2k}\left({\mu^2\over w^2}\right)\to(2k)!
\left(2-4^{1-k}\right)\zeta(2k)
\label{even}
\end{equation}
on even powers. If the imaginary part is fairly well behaved, as at large
$N_f$, it is unlikely that its euclideanization will be so.
Indeed, the factorial growth of the r.h.s. of Eq.(\ref{even})
should restore the factorial growth of the 
perturbative series for the euclidean quantity $\widetilde{D}_S$, expected
from general grounds of quantum field theory.  
The large-$N_f$
analysis of Table~3 suggests that the imaginary part $R_S$ is rather
well behaved for $n<7$ loops, with the far worse behaviour of $\widetilde{D}_S$
(and thus $\widetilde{\Delta}_n$)
resulting from its renormalon at $\de=1$, which was suppressed
by the sine function in~(\ref{Dpn}).

Next we consider the more prudent euclideanization~(\ref{Db}),
with a perturbation series
\begin{eqnarray}
\overline{D}_S(Q^2)&=&3[m(Q^2)]^2\left(1+\sum_{n>0}\overline{d}_n
\left({\alpha_s(Q^2)\over\pi}\right)^n\right)\label{dbn}\\
\overline{d}_n&=&d_n-2\gamma_{n-1}-
\sum_{k=1}^{n-1}(k\beta_{n-k-1}+2\gamma_{n-k-1})d_k
\label{dbk}
\end{eqnarray}
The large-$N_f$
results for $\overline{D}_S$ are obtained from
\begin{eqnarray}
\overline{d}_{n-1}&=&(-N_f/6)^{n-2}
\left(A_n+\overline{\Delta}_n+O(1/N_f)\right)
\label{dbl}\\
\overline{\Delta}_n&=&\widetilde{\Delta}_n-(n-2)\widetilde{\Delta}_{n-1}
\label{Dbr}
\end{eqnarray}
at $n>2$ loops,
with the $\de=1$ renormalon removed by the combination~(\ref{Dbr}).
At 2 loops, we have $\overline{\Delta}_2=\widetilde{\Delta}_n-2=16/3$.
Table~3 shows that the factorial growth in $\overline{D}_S$ at large $N_f$
is milder than in $\widetilde{D}_S$, but more severe
than in the imaginary part $R_S$.

From~(\ref{RD},\ref{dbk}) we obtain the first 5 terms
in the expansion
\begin{eqnarray}
R_D(Q^2)&=&\sum_{n\ge0}r_n\left({\alpha_s(Q^2)\over\pi}\right)^{n+1}
\label{RDQ}\\
r_0&=&\gamma_0=1\label{r0}\\
r_1&=&\gamma_1+\frac{1}{2}\beta_0d_1
\label{r1}\\
r_2&=&\gamma_2+\beta_0d_2
-\frac{1}{2}\beta_0d_1^2+\frac{1}{2}\beta_1d_1
\label{r2}\\
r_3&=&\gamma_3+\frac{3}{2}\beta_0d_3
+\frac{1}{2}\beta_0d_1^3-
\frac{3}{2}\beta_0d_1d_2
+\beta_1d_2-\frac{1}{2}\beta_1d_1^2+\frac{1}{2}\beta_2d_1
\label{r3}\\
r_4&=&\gamma_4+2\beta_0d_4+2\beta_0d_1^2d_2-2\beta_0d_1d_3
-\frac{1}{2}\beta_0d_1^4-\beta_0d_2^2\nonumber\\&&{}
+\frac{3}{2}\beta_1d_3+\frac{1}{2}\beta_1d_1^3-\frac{3}{2}\beta_1d_1d_2
-\frac{1}{2}\beta_2d_1^2+\beta_2d_2+\frac{1}{2}\beta_3d_1
\label{r4}
\end{eqnarray}
where the 5-loop coefficient $d_4$ is unknown, while  the estimates
for the 5-loop anomalous quark mass dimension are  known from 
the results of application of the Pad\'e resummation method 
\cite{EJJKS} ( note, however, that the analytical calculation 
of order $O(1/N_f^2)$-corrections to $\gamma_m(\alpha_s)$ \cite{JG2}
indicate that the latter ones should be refined). 
At large $N_f$ we simply obtain
\begin{equation}
r_n=(-N_f/6)^n\left({n\over2}\widetilde{\Delta}_{n+1}+O(1/N_f)\right)
\label{rl}
\end{equation}
for $n>0$, with no contribution from the anomalous quark-mass dimension
beyond the one-loop result $r_0=\gamma_0=1$.
It follow that at large-$N_f$ the perturbation series for $R_D$
explodes as violently as that for $\widetilde{D}_S$.

\section{Subtleties of the naive nonabelianization}

In the previous section we mainly concentrated on
the analysis  of large $N_f$ perturbative results, 
for the different theoretical quantities, related 
to the correlator of quark scalar currents. 
However, as was already 
explained in Sec.2.7, it is of definite interest to study the truncated 
and Borel-resummed perturbative series within the framework of 
the NNA  Ansatz, which is postulated 
by applying the    
substitution 
$N_f\to {N_f-11N_c/2}$=$-6\beta_0$ (where $\beta_0$ is the first coefficient
of the QCD $\beta$-function, defined by Eq.(1)), supplemented by retaining  
the leading terms in powers of $\beta_0$ in the reorganized perturbative 
series. This procedure enables one to
transform large-$N_f$ results, which are related to QED 
(note that in QED $\beta_0$ is 
proportional to $N_f$), to the nonabelian case of QCD.

In this section we shall study a number of theoretical issues 
related to the application of the NNA approximation in the scalar 
channel. In particular, we shall concentrate on 
obtaining  estimates of uncalculated  higher-order terms in 
perturbative series for  quantities, related to 
the correlator of quark scalar currents both in the euclidean and minkowskian 
regions. We shall also formulate different  procedures for the resummation 
of the large  minkowskian $\pi^2$-terms, within the framework of the NNA 
approach.

\subsection{Estimates of the higher order corrections in the 
$\overline{\rm MS}$-scheme.}

We begin by considering the expressions 
\begin{eqnarray}
d_n^{\rm NNA}&=&\beta_0^{n-1}\widetilde{\Delta}_{n+1}\label{dnna}\\
s_n^{\rm NNA}&=&\beta_0^{n-1}\Delta_{n+1}\label{snna}\\
\overline{d}_n^{\rm NNA}&=&\beta_0^{n-1}\overline{\Delta}_{n+1}
\label{dbnna}
\end{eqnarray}
for the coefficients $d_n$, $s_n$ and $\overline{d}_n$
in $\widetilde{D}_S$, $R_S$ and $\overline{D}_S$ at $n+1$ loops
in the $\overline{\rm MS}$-scheme.
These are obtained by the naive nonabelianization 
of the terms in~(\ref{dtl},\ref{sl},\ref{dbl})
that do not involve the large-$N_f$
anomalous quark-mass dimension contribution $A_n$,
which is known to be unamenable to naive nonabelianization in general 
(for detailed discussions, related to  the deep-inelastic scattering 
anomalous dimensions see the works of  
Ref.\cite{SM}),
and in any case is small at 5 loops, since it falls off like $(2/5)^n$.

Table~4 shows that all three NNA estimators give the correct sign 
and order of magnitude at
3 and 4 loops. At 4 loops, the success of $\overline{d}_3^{\rm NNA}$ is rather
remarkable, since we are using only the $(N_f-33/2)^2$ approximation
to a quadratic
to estimate the full result. In particular, at $N_f=5$, as in Higgs decay,
$\overline{d}_3$ differs from the NNA estimate by only 8\%.
It is significant that this success
of NNA occurs in the safer euclideanization $\overline{D}_S$,
which includes neither the $\pi^2$ terms of $R_S$ nor the spurious
renormalon of $\widetilde{D}_S$. Accordingly we take
\begin{equation}
\overline{d}_4^{\rm NNA}
=\beta_0^3\left({17597\over324}+{20\over3}\zeta(3)-60\zeta(5)\right)
\label{best}
\end{equation}
as our favoured NNA estimator.

Note, that even if we choose an overall factor of two as the conservative uncertainty 
in the estimating power of the NNA procedure (which is motivated by 
inspecting the related numbers of Table 4 for $N_f=3$), 
we arrive to the conclusion 
that the 5-loop perturbative approximation for $\overline{D}_S$ is really 
well-behaved. Indeed, taking $N_f=3$ and $\alpha_s\approx$ 0.3 we obtain the 
following  series 
\begin{eqnarray}
\overline{D}_S &=& 1 + 3.67\bigg( \frac{\alpha_s}{\pi}\bigg) + 
14.17 \bigg(\frac{\alpha_s}{\pi}\bigg)^2 + 
77.36 \bigg(\frac{\alpha_s}{\pi}\bigg)^3 + 
2\times 1.26 \bigg(\frac{\alpha_s}{\pi}\bigg)^4 \\ \nonumber 
&=& 1 + 0.350 + 0.129 + 0.067 + 0.0002
\end{eqnarray}
with a  rather small 5-loop term. 
Thus, the 4-loop extractions of the running mass $m_s$ of Ref.\cite{CPS}, 
which is based  on the consideration of the  $\overline{D}_S$-function, 
indeed contains rather small perturbative QCD uncertainties 
due to the truncation of the corresponding series at the 4-loop 
level. 

Let us now turn to the study of the NNA predictions for 
the coefficients of the perturbative series for $R_S$ in the minkowskian 
region. As in the case of $\overline{D}_S$ and $\widetilde{D}_S$-functions, 
this procedure gives the correct sign and order of magnitude at the 
3- and 4-loop levels (see Table 4). Taking into account 
the large $N_f$-result for $\Delta_5$ (see Eq.(57)) we get the following 
NNA prediction for the 5-loop term in $R_S$
\begin{equation}
s_4^{\rm NNA} = \beta_0^3\left(\frac{64877}{324}-\frac{100}{3}\zeta(3)
-60\zeta(5)-\frac{275}{18}\pi^2+4\zeta(3)\pi^2+\frac{1}{10}\pi^4\right)
\end{equation}
which gives small and positive numbers
\begin{equation}
s_4^{\rm NNA}(N_f=3)\approx49\,;\quad
s_4^{\rm NNA}(N_f=4)\approx39\,;\quad
s_4^{\rm NNA}(N_f=5)\approx31
\label{NNA}
\end{equation} 
Following now the conservative 
pattern of fixing the uncertainty of the NNA approximation of 
$\overline{d}_n$-terms by an overall factor of 2, we present  the NNA-inspired 
estimates of $s_4$-term in the following form: $s_4\approx 2s_4^{\rm NNA}$.
Keeping this in mind, we arrive at the following numerical estimates
of $s_4$ for different numbers of $N_f$: 
\begin{equation}
s_4(N_f=3)\sim 98\,;\quad
s_4(N_f=4)\sim 78\,;\quad
s_4(N_f=5)\sim 62
\label{NNA2}
\end{equation}
For $N_f=5$ these estimates are quite in accord with the result 
of applying the [2/2] asymptotic Pad\'e-approximation method, 
namely 
$s_4^{APAP}(N_f=5)\approx 67$ \cite{HPade}. 
However, for $N_f=3$ the Pad\'e estimate of Ref.\cite{HPade},
namely $s_4^{APAP}(N_f=3)\approx 251$, 
is over 2.6 times larger than the related NNA inspired 
estimate of Eq.(\ref{NNA2}).

Note, that contrary to what was found in the application of the 
Pad\'e-resummation method to the  $e^+e^-$ annihilation $R$-ratio \cite{SEK}, 
the variant of asymptotic Pad\'e estimates used in Ref.\cite{HPade},
which is performed in the minkowskian region directly,  
does not allow one to separate the effects 
of analytical continuation proportional to $\pi^2$. In the NNA approach the  
contributions of the $\pi^2$-effects leading in ${\beta}_{0}$ are taken into account explicitly 
and can be resummed to all-orders of perturbation theory. We shall consider 
this technical problem in Sec.3.3. Another  observation is that 
the estimates of Eq.(\ref{NNA2}) differ in both sign and order of magnitude 
from the ones obtained 
in Ref.\cite{CKS} using a  variant of the effective-charges 
procedure  \cite{KS,KS2}. In Sec.3.6 we shall return to 
more detailed considerations of the problems
 related 
to the application of the effective charges method in the scalar channel. 

We conclude this subsection by demonstrating the behaviour of the 
$\overline{\rm MS}$- perturbative series for the  physical quantity 
$R_S$ of Eq.(71) in the cases of $N_f=3$ and $N_f=5$, which are relevant 
to the spectral function of the QCD sum rules and the hadronic decay width 
of the Higgs boson. Taking $\alpha_s\approx 0.3$ in the first case 
and $\alpha_s\approx 0.114$ in the latter one we have 
\begin{eqnarray}
N_f=3~:~R_S &\sim& 1+5.667\bigg(\frac{\alpha_s}{\pi}\bigg)+
31.864\bigg(\frac{\alpha_s}{\pi}\bigg)^2+
89.156\bigg(\frac{\alpha_s}{\pi}\bigg)^3+98\bigg(\frac{\alpha_s}{\pi}\bigg)^4
\\ \nonumber 
&=& 1+ 0.541 + 0.291 + 0.078 + 0.008 
\label{RS3}
\end{eqnarray} 
\begin{eqnarray}
N_f=5~:~R_S &\sim& 1+5.667\bigg(\frac{\alpha_s}{\pi}\bigg)+
29.147\bigg(\frac{\alpha_s}{\pi}\bigg)^2+
41.758\bigg(\frac{\alpha_s}{\pi}\bigg)^3+62\bigg(\frac{\alpha_s}{\pi}\bigg)^4
\\ \nonumber 
&=& 1+ 0.206 + 0.0384 + 0.0021 + 0.00014 
\label{RS5}
\end{eqnarray} 
One can see, that in both cases the perturbative series are rather 
well behaved and that the NNA-inspired estimates of 5-loop terms are 
over 10 times 
smaller than the 4-loop ones explicitly calculated in Ref.\cite{KC} .
In view of this we conclude that the 4-loop phenomenological studies, 
based on the 4-loop series of  (110,111) are in rather good shape, and 
that the manifestation of the asymptotic growth of these 
perturbative series is postponed. Indeed, in accordance with the results 
of Tables 2,3 this feature can manifest itself starting from $n=7$ 
 loops. In its turn, this means  that in the process of 
concrete phenomenological applications 
of the perturbative results for $R_S$ in the energy region where 
$\alpha_s\leq 0.3$ one can restrict oneself to a consideration of the partial sums 
of the truncated perturbative series with $n\leq 6$  loops, 
estimating roughly the remaining perturbative uncertainty in the 
$\overline{\rm MS}$-scheme by the value of the smallest term taken into 
account. More detailed numerical studies are performed in Sec. 3.5.

\subsection{  Large $N_f$ versus $N_c$ -theoretical motivation for NNA}

It is interesting to consider the motivation for the NNA approximation
from the theoretical point of view. In
the vector case the NNA term can be proved to have some very special
properties by analyzing the operators that build the leading UV
renormalon singularity \cite{r19,r20}. It is convenient to consider a `planar
approximation' where at each order in the $N_f$ expansion only terms leading
in $N_c$ are retained. In this way one obtains an expansion
in multinomials of $N_f$ and $N_c$. So for the Adler-function coefficients in 
the
vector case one can write,
after extracting an overall factor of $(3/4){C_F}$ 
\cite{Chris,r22},
\begin{equation}
{d_n}={d}_{n}^{[n]}{N}_{f}^{n}+{d}_{n}^{[n-1]}{N}_{f}^{n-1}{N_c}+
{d}_{n}^{[n-2]}{N}_{f}^{n-2}{N_c}^{2}
+{\ldots}+{d}_{n}^{[1]}{N_f}{N_c}^{n-1}+{d}_{n}^{[0]}{N_c}^{n}\;,
\end{equation}
so that the large-$N_f$ expansion runs from left-to-right, and the large-$N_c$
from right-to-left. One can now formulate two versions of NNA.
The standard one is derived by replacing $N_f$ by $(11N_c-12{\beta}_{0})/2$ 
to arrive at
\begin{equation}
{d}_{n}={d}_{n}^{(n)}{{\beta}_{0}}^{n}+
{d}_{n}^{(n-1)}{{\beta}_{0}}^{n-1}{N_c}+{d}_{n}^{(n-2)}{{\beta}_{0}}^{n-2}
{N_c}^{2}
+{\ldots}+{d}_{n}^{(1)}{{\beta}_{0}}{N_c}^{n-1}+{d}_{n}^{(0)}{N_c}^{n}\;.
\end{equation}
One can, however , define a `dual NNA' by replacing $N_c$ by 
$(12{\beta}_{0}+2{N_f})/11$, to
obtain an expansion in ${\beta}_{0}$ with different coefficients. 
The standard NNA is
exact in the large-$N_f$ limit, and the `dual NNA' is exact in the
large-$N_c$ limit. Of course it is the standard NNA that is of practical use
since we have all-orders large-$N_f$ results. If one extracts the NNA term 
${d}_{n}^{NNA}$ of the standard expansion and re-expands it one can obtain an
expansion in $N_f$ and $N_c$  with coefficients 
${\tilde{d}}_{n}^{[n-r]}$.
By construction the leading-$N_f$ term is reproduced, so 
${d}_{n}^{[n]}={\tilde{d}}_{n}^{[n]}$,
but the sub-leading terms will not be reproduced. 
Nonetheless by making use of
the operator analysis of \cite{r19} one can show that \cite{r20},
\begin{equation}
{d}_{n}^{[n-r]}{\approx}{\tilde{d}}_{n}^{[n-r]}[1+O(1/n)]\;,
\end{equation}
so that for fixed-$r$ and large orders of perturbation theory the re-expansion
of the NNA term approximates the sub-leading in $N_f$ terms with
asymptotic accuracy O($1/n$). For the dual NNA term one can prove an
exactly similar result, where sub-leading in $N_c$ terms are reproduced to
asymptotic accuracy O($1/n$) on re-expansion of the dual NNA term \cite{r20}.
Such weak asymptotic results about the NNA terms will hold provided
that in the large-$N_f$, and large-$N_c$ limits the leading UV renormalon 
asymptotics is controlled by a {\it single} operator contribution. This is
the case for the vector Adler function of Eq.(60). In planar approximation 
there are two
relevant four-fermion operators, ${\cal{O}}_{+}$ and ${\cal{O}}_{-}$ of 
\cite{r20},
 but ${\cal{O}}_{-}$ is scalar after Fierzing and
decouples. 
These operators are defined as \cite{r20}
\[
{\cal{O}}_{\pm} = {\cal{O}}_V \pm {\cal{O}}_A 
\] 
\[
{\cal{O}}_V = 
\left(\overline{\psi}\gamma_{\mu}T^{A}\psi\right)
\left(\overline{\psi}\gamma^{\mu}T^{A}\psi\right)~~~,~~~ 
{\cal{O}}_A = 
\left(\overline{\psi}\gamma_{\mu}\gamma_5T^{A}\psi\right)
\left(\overline{\psi}\gamma^{\mu}\gamma_5T^{A}\psi\right) \\ 
\]
where $T^A$ denotes the colour matrices.
The remaining four-fermion operator ${\cal{O}}_{+}$ gives the 
leading asymptotics
in the large-$N_c$ limit , and in the large-$N_f$ limit the operator 
corresponding
to the single renormalon chains involved in NNA  
(${\cal{O}}_{1}$ of \cite{r20}) dominates the asymptotics 
\cite{r19}.
This operator is defined as \cite{r20}
\[
{\cal{O}}_1 = (1/g^4) \partial_{\nu} F^{\nu\mu}\partial^{\rho}F_{\rho\mu}~~~.
\]

 Let us see how these weak asymptotic
results work by re-expanding the vector $d_1$,$d_2$, explicitly. We shall
denote the dual NNA term by ${d}_{n}^{NNA*}$, the ${\overline{\rm MS}}$ scheme 
with ${\mu^2}=Q^2 $ is
assumed,
\begin{eqnarray}
{d_1}&=&-0.115{N_f}+0.655N_c
\nonumber \\
{d}_{1}^{NNA}&=&-0.115{N_f}+0.643N_c
\nonumber \\
{d}_{1}^{NNA*}&=&-0.119{N_f}+0.655N_c\;,
\end{eqnarray}
and 
\begin{eqnarray}
{d_2}&=&0.086{N}_{f}^{2}-1.40{N_f}{N_c}+2.10{N_c}^{2}
\nonumber \\
{d}_{2}^{NNA}&=&0.086{N}_{f}^{2}-0.948{N_f}{N_c}+2.61{N_c}^{2}
\nonumber \\
{d}_{2}^{NNA*}&=&0.069{N}_{f}^{2}-0.763{N_f}{N_c}+2.10{N_c}^{2}\;.
\end{eqnarray}
So we see that the weak asymptotic property holds, and the two
versions of NNA give surprisingly good approximations for
the sub-leading in $N_f$ and $N_c$ coefficients, given that this is only
supposed to be an asymptotic result.\\

Now let us repeat this analysis  
for the scalar $\widetilde{D}_S$-function of Eq.(67). We find
\begin{eqnarray}
{d_2}&=&-1.91{N_f}+17.19{N_c}
\nonumber \\
{d}_{2}^{NNA}&=&-1.91{N_f}+10.49{N_c}
\nonumber \\
{d}_{2}^{NNA*}&=&-3.13{N_f}+17.19{N_c}\;,
\end{eqnarray}
and
\begin{eqnarray}
{d_3}&=&0.93{N}_{f}^{2}-21.25{N_f}{N_c}+72.08{N_c}^{2}
\nonumber \\
{d}_{3}^{NNA}&=&0.93{N}_{f}^{2}-10.22{N_f}{N_c}+28.10{N_c}^{2}
\nonumber \\
{d}_{3}^{NNA*}&=&2.38{N}_{f}^{2}-26.21{N_f}{N_c}+72.08{N_c}^{2}
\end{eqnarray}
So we see that whilst the dual NNA term works reasonably well,
the standard NNA in which we are interested yields the
subleading in $N_f$ coefficients with correct sign, but significantly
reduced accuracy.
For the coefficients of more physically-interesting 
quantity $R_S$ of Eq.(71), which as we saw from the analysis of Sec.2.8,
is perturbatively better behaved, we find 
\begin{eqnarray}
{s_2}&=&-1.36{N_f}+11.98{N_c}
\nonumber \\
{s}_{2}^{NNA}&=&-1.36{N_f}+7.47{N_c}
\nonumber \\
{s}_{2}^{NNA*}&=&-2.18{N_f}+11.98{N_c}\;,
\end{eqnarray}
and
\begin{eqnarray}
{s_3}&=&0.26{N}_{f}^{2}-8.59{N_f}{N_c}+18.24{N_c}^{2}
\nonumber \\
{s}_{3}^{NNA}&=&0.26{N}_{f}^{2}-2.85{N_f}{N_c}+7.83{N_c}^{2}
\nonumber \\
{s}_{3}^{NNA*}&=&0.60{N}_{f}^{2}-6.63{N_f}{N_c}+18.24{N_c}^{2}\;.
\end{eqnarray}
We see that, again, the dual NNA term works quite well, but
the standard version performs 
less satisfactorily.
As in the vector case, this can be understood in terms of the operators
involved \cite{r23}. In the scalar case it is the four-fermion operator ${\cal{O}}_{+}$
which previously dominated the large-$N_c$ asymptotics which is vector
after Fierzing and decouples. The remaining four-fermion operator ${\cal{O}}_{-}$ will
dominate the large-$N_c$ asymptotics, underwriting the success of the
dual NNA. In the large-$N_f$ limit, however, it turns out that ${\cal{O}}_{-}$
 {\it and} the one-chain operator in the scalar case are
{\it both} involved in determining the asymptotics, and so standard
NNA will not satisfy the weak asymptotic result that sub-leading in
$N_f$ terms are reproduced.
 Whilst the special
property of NNA  which holds for the vector case will
not be true for the scalar, the numerical accuracy of the
approximation is not so bad (see Table 4) 
despite the less satisfactory performance evident from 
the results of 
Eqs.(117)-(120).

\subsection{Analytic continuation of fractional powers of ${\alpha}_{s}$} 

In this  section we shall study the analytical continuation of the euclidean
construct ${\widetilde{D}}_{S}(Q^2)$ introduced in (64). This is related to the
quantity $R_S$ by an analytical continuation to the
minkowskian region. The continuation is essentially the same as
that involved in the vector case for the analytical continuation of the
Adler $D$-function to the QCD $R$-ratio. 
The effects of analytical continuation 
have been much studied \cite{r0}-\cite{r11} in
attempts to improve the convergence of the QCD  perturbation series 
 by resumming an infinite subset of analytical continuation terms
at each order in perturbation theory.
Such resummation may be accomplished
conveniently by representing the continuation as a contour integral around
a circle in the complex $-{Q}^{2}$ plane
(for one of the first discussion of this realization of the 
contour-improved technique see Ref.\cite{r5}).
 One can then perform the contour integration
numerically , at some given order of perturbation theory. In the process one
resums an infinite subset of potentially large analytical continuation
terms involving powers of ${\pi}^{2}$, which arise in the
running of the coupling around the circular contour. Such an expansion is
termed ``contour-improved'' .
For the case of a 
one-loop
coupling an explicit closed-form result can be given for 
the contour integral.
We would like to generalize these results to the present case where
the mass anomalous dimension gives rise  to fractional powers of 
${\alpha}_s$ (recent analogous independent considerations 
were given in Ref.\cite{r12}).
Using the NNA all-orders results for ${\widetilde{D}}_{S}$ and
$R_S$ we shall then perform various numerical studies on the performance
of fixed-order perturbation theory, and its ``contour-improved'' version.

The analytical continuation between $\widetilde{D}$ and $R$ of (67) and (71) 
can 
be written in
the form

\begin{equation}
 R_S(w^2)=
\frac{1}{2{\pi}}
{\int_{-{\pi}}^{\pi}}{d{\theta}}\;3 {[{m}({e}^{i{\theta}}{w^2})]}^{2}
\left(1+{\sum_{n>0}}{d_n}{\left(\frac{{\alpha}_{s}({e}^{i{\theta}}{w^2})}{\pi}
\right)}^{n}\right)\;,
\end{equation}
involving a contour integral around a circle in the complex $w^2=-{Q^2}$ 
plane, 
as mentioned above.

We can relate the running mass $m(Q^2)$ to the 
renormalization scheme invariant mass 
$\hat{m}$ as follows (see e.g. \cite{r13,Becchi}):
\begin{equation}
m(Q^2)=\hat{m}~ {\exp}\left[-\int^{\alpha_s(Q^2)}\frac{\gamma_m(x)}{\beta(x)}dx
+\frac{\gamma_0}{\beta_0}ln(2\beta_0)\right]\;,
\label{mass}
\end{equation} 
where the second term in the exponent is the commonly used normalization
of the definition of the invariant mass $\hat{m}$. 
Since we shall be working within the NNA procedure, 
we shall
 set ${\gamma}_{i}=0, (i>0)$ and $\beta_{i}=0 (i>0)$.
In this approximation one has 
 \begin{equation}
{[m(Q^2)]}^{2}={\hat{m}}^{2}{(2{\beta}_{0})}^{2{\gamma}_{0}/{\beta}_{0}}
{\left(\frac{{\alpha}_{s}(Q^2)}{\pi}\right)
 }^{2{\gamma}_{0}/{\beta}_{0}}\;.
\end{equation}
Inserting this expression for the running mass into (121) one arrives at
\begin{equation}
R_S(w^2)=3{\hat{m}}^{2}{(2{\beta}_{0})}^{2{\gamma}_{0}/{\beta}_{0}}
 \frac{1}{2{\pi}}{\int_{-\pi}^{\pi}}{d{\theta}}
{\left(\frac{{\alpha}_{s}({e}^{i{\theta}}w^2)}
 {\pi}\right)}
 ^{2{\gamma}_{0}/{\beta}_{0}}\left(1+{\sum_{n>0}}{d_n}
{\left(\frac{{\alpha}_{s}({e}^{i{\theta}}w^2)}{\pi}
 \right)}^{n}\right)\;.
 \end{equation}
The contour-improved expansion is obtained by performing the integration 
term-by-term.
For a one-loop beta-function, appropriate for the NNA approximation, one has
 \begin{equation}
 {\alpha}_{s}({e}^{i{\theta}}w^2)=\frac{{\alpha}_{s}(w^2)}
{[1+i{\beta}_{0}{\theta}{\alpha}_{s}(w^2)/{\pi}]}\;,
 \end{equation}
 and so ${d_n}$ in the ``contour-improved" NNA  expansion will be multiplied by
 \begin{equation}
 \frac{1}{2\pi}{\int_{-\pi}^{\pi}}\;{d{\theta}} 
\frac{{[{\alpha}_{s}(w^2)/\pi]}^{2{\gamma}_{0}/{\beta}_{0}+n}}
 {{[1+i{\beta}_{0}{\theta}{\alpha}_{s}({w^2})/{\pi}]}^{2{\gamma}_{0}/{\beta}_
{0}+n}}
 {\equiv} {\left(\frac{{\alpha}_{s}(w^2)}{\pi}\right)}^{2{\gamma}_{0}/{\beta}_
{0}}{A_n}({\alpha}_{s}(w^2))\;.
 \end{equation}
 The function ${A_n}({\alpha}_{s})$ is given in closed form by
 \begin{equation}
 {A_n}({\alpha}_{s})=\frac{1}{{\beta}_{0}{\delta}_{n}{\pi}}
{\left(1+{\beta}_{0}^{2}{{\alpha}_{s}^{2}}\right)}
^{{\delta}_{n}/2}
 {\left(\frac{{\alpha}_{s}}{\pi}\right)}^{n-1}{\sin}\left[{\delta}_{n}
{\arctan}\left({\beta}_{0}
 {\alpha}_{s}\right)\right]\;,
 \end{equation}
 where ${\delta}_{n}{\equiv}1-n-2{\gamma}_{0}/{\beta}_{0}$. 
For ${\delta}_{n}\rightarrow{0}$ this
reproduces the well-known factor 
$(1/{\pi}{\beta}_{0}){\arctan}({\beta}_{0}{\alpha}_{s})$
 obtained by resumming all the analytical continuation terms only involving 
${\beta}_{0}$
 for the ${e}^{+}{e}^{-}$  $R$-ratio, 
while for $n=0$ the expansion of the 
r.h.s. of Eq.(127) to first order in 
$\alpha_s=\pi/(\beta_0ln(w^2/\Lambda^2))$ coincides with the result previously 
obtained in \cite{Gor1}. 
 Finally we can write down two expansions for
 $R$ with NNA,
 \begin{equation}
 R_S=3{\hat{m}}^{2}{(2{\beta}_{0})}^{2{\gamma}_{0}/{\beta}_{0}}
{\left(\frac{{\alpha}_{s}(w^2)}{\pi}\right)}
 ^{2{\gamma}_{0}/{\beta}_{0}}
 \left(1+{\sum_{n>0}}{s_{n}^{NNA}}
 {\left(\frac{{\alpha}_{s}(w^2)}{\pi}\right)}^{n}\right)\;,
 \end{equation}
 or, alternatively, the ``contour-improved'' NNA expansion,
\begin{equation}
 R_S=3{\hat{m}}^{2}{(2{\beta}_{0})}^{2{\gamma}_{0}/{\beta}_{0}}
{\left(\frac{{\alpha}_{s}(Q)}{\pi}\right)}^
 {2{\gamma}_{0}/{\beta}_{0}}
 \left({A}_{0}^{NNA}({\alpha}_{s}(w^2))+{\sum_{n>0}}{d_{n}^{NNA}}
 {A}_{n}^{NNA}({\alpha}_{s}(w^2))\right)\;.
 \end{equation}
 The $A_{n}^{NNA}$ for $n>1$ are defined from (127) on setting 
${\delta}_{n}=1-n$. For $n=0$
 one needs to be careful. 
The NNA terms are of the form ${\beta}_{0}^{i-1}{\alpha}_{s}^{i}$,
 with $n=0$ one needs to isolate the terms linear in ${\gamma}_{0}$ that 
arise on expanding (127) in powers of
 ${\alpha}_{s}$.
 One finds 
 \begin{equation}
 {A}_{0}^{NNA}({\alpha}_{s})=1-\frac{{\gamma}_{0}}{{\beta}_{0}}{\ln}
(1+{{\beta}_{0}}^{2}{\alpha_{s}}^{2})
 -\frac{2{\gamma}_{0}}{{{\beta}_{0}}^{2}{\alpha}_{s}}{\arctan}({\beta}_{0}
{\alpha}_{s})+\frac
 {2{\gamma}_{0}}{{\beta}_{0}}\;.
 \end{equation}
Note that this term contains {\it all} contributions depending 
on the anomalous
dimension ${\gamma}_{0}$. The remaining ${A}_{n}^{NNA}$ for $n>0$ are 
precisely 
the same as the functions which arise in the case of the ${e}^{+}{e}^{-}$
$R$-ratio.
The $n=1$ case corresponds to ${\delta}_{n}\rightarrow{0}$ and so one has the
well-known ${\arctan}$ alluded to earlier,
 \begin{equation}
 {A}_{1}^{NNA}({\alpha}_{s})=\frac{1}{{\pi}{\beta}_{0}}{\arctan}({\beta}_{0}
{\alpha}_{s})\;.
 \end{equation}
 \\
 For $n>1$ the ${A}_{n}^{NNA}$ are in fact simple rational functions of
 ${\alpha}_{s}/{\pi}$. One has, for instance,
 \begin{eqnarray}
 {A}_{2}^{NNA}({\alpha}_{s})&=&\frac{{({\alpha}_{s}/{\pi})}^{2}}
{1+{\beta}_{0}^{2}{\alpha}_{s}^{2}}
 \nonumber \\
 {A}_{3}^{NNA}({\alpha}_{s})&=&\frac{{({\alpha}_{s}/{\pi})}^{3}}
{{(1+{\beta}_{0}^{2}{\alpha}_{s}^{2})}^{2}}
\nonumber \\
{A}_{4}^{NNA}({\alpha}_{s})&=&\left({\left(\frac{{\alpha}_{s}}
{\pi}\right)}^{4}-
\frac{{\pi}^{2}{\beta}_{0}^{2}}{3}{\left(\frac{{\alpha}_{s}}{\pi}
\right)}^{6}\right)/
{(1+{\beta}_{0}^{2}{\alpha}_{s}^{2})}^{3}\;.
\end{eqnarray}

\subsection{Scheme dependence of NNA results}
 Before proceeding to some numerical studies we need to confront one 
further important subtlety. As
 we have defined them the NNA expansions in (128) and (129) are 
scheme-dependent. 
Of course, we expect the partial sums to be scheme-dependent. The problem
is that the 
 {\it all-orders} sum for $R_S$ will depend on our choice of 
renormalization scale. Since $R_S$ is
 a physical quantity this is clearly undesirable.
  Following
 other similar analyses for the vector correlator \cite{r7,r22,r24, Rid} 
we  will use in the next section  a Borel sum
of the  divergent series to define the all-orders sum, using regulation to 
cope with the
IR renormalon contributions. 
The all-orders sum so defined  can then be compared with
fixed-order perturbation theory partial sums to obtain an estimate of 
the likely effect
of uncalculated higher-order corrections. The problem is that 
the all-orders (Borel) sum of the
series in (128) combined with the fractional power 
of ${({\alpha}_{s}/{\pi})}^{2{\gamma}_{0}/{\beta}_{0}}$
depends on the renormalization scale used for ${\alpha}_{s}$. 
The difficulty is the
fractional power of ${\alpha}_{s}$ involving $1/{\beta}_{0}$. 
 For illustrative purposes
suppose that we used
 the so-called $V$-scheme 
(see e.g. Ref.\cite{V})
corresponding to $\overline{\rm MS}$ with ${\mu^2}={e}^{-5/3}w^2$ rather than
 ${\mu^2}=w^2$. 
 Writing ${\alpha}_{s}^{V}$ and ${\alpha}_{s}^{\overline{\rm MS}}$ for the 
two scale choices, we
 have,  assuming a one-loop beta-function,
 \begin{eqnarray}
&&{\left(\frac{{\alpha}_{s}^{\overline{MS}}}{\pi}\right)}^
{2{\gamma}_{0}/{\beta}_{0}}=
  {\left(\frac{{\alpha}_{s}^{V}}{\pi}\right)}^{2{\gamma}_{0}/{\beta}_{0}}
  {\left[1+\frac{5}{3}{\beta}_{0}\frac{{\alpha}_{s}^{V}}{\pi}\right]}^
{-2{\gamma}_{0}/{\beta}_{0}}
  \nonumber \\
  &=&{\left(\frac{{\alpha}_{s}^{V}}{\pi}\right)}^{2{\gamma}_{0}/{\beta}_{0}}
  \left[1-\frac{10}{3}{\gamma}_{0}\left(\frac{{\alpha}_{s}^{V}}{\pi}\right)+
  \frac{2{\gamma}_{0}}{\beta_{0}}\left(\frac{2{\gamma}_{0}}
{\beta_{0}}+1\right)\frac{25}{18}{{\beta}_{0}}^{2}
  {\left(\frac{{\alpha}_{s}^{V}}{\pi}\right)}^{2}+{\ldots}\right]\;.
  \end{eqnarray}
  Recalling that the NNA terms have the structure 
${\beta}_{0}^{i-1}{\alpha}_{s}^{i}$ it
is clear that only the terms linear in ${\gamma}_{0}$ in the 
expansion in the second line,
  appear in NNA. Thus, since not all terms involved in the change of 
scheme are retained,
  the resummed NNA expansions will be scheme-dependent. The resolution of 
this problem is to avoid
  powers of ${\alpha}_{s}$ involving $1/{\beta}_{0}$. This can be 
accomplished by identifying an effective charge 
$\hat{R}$ related to $R$ by \cite{Grunberg1,r18},
  \begin{equation}
  R_S=3{\hat{m}}^{2}{[2{\beta}_{0}{\hat{R}}]}^{2{\gamma}_{0}/{\beta}_{0}}\;.
  \end{equation}
  ${\hat{R}}$ will have the perturbative expansion
  \begin{equation}
  {\hat{R}}=\left(\frac{{\alpha}_{s}(w^2)}{\pi}\right)\left(1+{\sum_{n>0}
}\;{\hat{s}}_{n}
  {\left(\frac{{\alpha}_{s}(w^2)}{\pi}\right)}^{n}\right)\;.
  \end{equation}
  Since this only involves integer powers of ${\alpha}_{s}$ 
all terms involved in a change of scheme at the one-loop level now 
contribute to the NNA
result, and the resummed NNA
  expansion will be scheme-independent.
\begin{equation}
  {R}_{S}=3{\hat{m}}^{2}{(2{\beta}_{0})}^{2{\gamma}_{0}/{\beta}_{0}}
{\left[\left(\frac{{\alpha}_{s}(w^2)}{\pi}\right)
  \left(1+{\sum_{n>0}
  }\;{\hat{s}}^{NNA}_{n}
{\left(\frac{{\alpha}_{s}(w^2)}{\pi}\right)}^{n}\right)\right]}
  ^{2{\gamma}_{0}/{\beta}_{0}}\;,
 \end{equation}  
is a reformulation of NNA for $R_S$ in which the resummed series is scheme-independent.
The unsatisfactory ``scheme-dependent'' version in (128) follows
  from it if only the terms linear in ${\gamma}_{0}$ 
are retained in expanding the series in
  ${\hat{s}}^{NNA}_i$.
 Writing 
$S^{2{\gamma}_{0}/{\beta}_{0}}={\exp}[(2{\gamma}_{0}/{\beta}_{0}){\ln}(S)]$
  and expanding the $\exp$ to O(${\gamma}_{0}$)  one arrives at
  \begin{equation}
  1+{\sum_{n>0}}\;{s}_{n}^{NNA}{\left(\frac{{\alpha}_{s}(w^2)}
{\pi}\right)}^{n}=
  1+\frac{2{\gamma}_{0}}{{\beta}_{0}}{\ln}\left[1+{\sum_{n>0}}\;
{\hat{s}}_{n}^{NNA}
  {\left(\frac{{\alpha}_{s}(w^2)}{\pi}\right)}^{n}\right]\;,
  \end{equation}
  which relates the two versions of NNA for $R_S$. Using this 
result one can rewrite
  the reformulated expansion of (136) in terms of the ${s}_{i}^{NNA}$,
  (128) is replaced by, 
\begin{equation}
  R_S=3{\hat{m}}^{2}{(2{\beta}_{0})}^{2{\gamma}_{0}/{\beta}_{0}}
{\left(\frac{{\alpha}_{s}(w^2)}{\pi}\right)}^
  {2{\gamma}_{0}/{\beta}_{0}}{\exp}\left[{\sum_{n>0}}\;{s}_{n}^{NNA}
{\left(\frac{{\alpha}_{s}(w^2)}{\pi}
  \right)}^{n}\right]\;.
  \end{equation}
  We can immediately write a contour-improved version which replaces (129),
  \begin{equation}
  R_S=3{\hat{m}}^{2}{(2{\beta}_{0})}^{2{\gamma}_{0}/{\beta}_{0}}
{\left(\frac{{\alpha}_{s}(w^2)}{\pi}\right)}^
  {2{\gamma}_{0}/{\beta}_{0}}{\exp}
\left[{A}_{0}^{NNA}({\alpha}_{s}(w^2))-1+{\sum_{n>0}}{d}_{n}^{NNA}
  {A}_{n}^{NNA}({\alpha}_{s}(w^2))\right]\;.
  \end{equation} 
\\
Using the Borel Sum to resum the series in the exponent in 
(138) and combining with the
fractional power of ${\alpha}_{s}$ we will now obtain an all-orders 
result for
$R_S$ which is independent of renormalization scale, as 
required since $R_S$ is a 
physical quantity. 

\subsection{Numerical studies on the convergence of the NNA results}
We shall now perform some numerical studies on the 
reformulated NNA expansions
of (138) and (139). We shall consider the partial sums 
${R}^{(n)}_{\overline{\rm MS}}$ and
${R}^{(n)}_{V}$ obtained by summing the series in the exponent in (138) 
up to and including
the ${s}_{n}^{NNA}$ term, in the $\overline{\rm MS}$ and $V$ schemes, 
respectively.
 The prefactor $3{\hat{m}}^{2}{(2{\beta}_{0})}^{2{\gamma}_{0}/{\beta}_{0}}$
is set to unity. 
We shall also consider the analogous partial sums of the contour-improved
expansion in (139), ${R}^{(n)CI}_{\overline{MS}}$ and ${R}^{(n)CI}_{V}$.
In Table 5 we begin by displaying the ${s}_{n}^{NNA}$ coefficients in the
${\overline{\rm MS}}$ scheme, and the $V$-scheme. 
We assume ${N_f}=5$ active flavours of
quarks. We shall use these two schemes to illustrate some of the scheme-dependence
subtleties discussed in the last Section. 
 In contrast to the ``plateau of tranquility'' evident in the limited
growth of the ${\overline{\rm MS}}$ scheme coefficients for $n<7$ ,
which was alluded to in
Sec. 2.9, we see that the corresponding $V$-scheme coefficients grow much more rapidly. 
Alternating
sign growth is evident even in low orders reflecting the asymptotic
alternating factorial growth contributed by the leading UV renormalon 
singularity
at ${\delta}=-1$ contained in the Borel transform ${G}_{-}({\delta})$ in (44).
In the ${\overline{\rm MS}}$ scheme this behaviour is temporarily screened in
low orders by the ${\exp}(5{\delta}/3)$ factor in (53), which is absent in the
$V$-scheme.
 In 
Table 6
we show the partial sums for the choice of coupling 
${\alpha}_{s}^{\overline{\rm MS}}=0.114$,
(${\alpha}_{s}^{V}=0.12895$) appropriate for Higgs width 
determination ($N_{f}=5$ is assumed).
As can be seen convergence is rapid for all the expansions. Of course, this is only temporary since
the series is asymptotic, and for sufficiently large orders the alternating factorial growth of
coefficients due to the leading ultra-violet renormalon will be evident. 
 The $V$-scheme leads to
slightly faster convergence than ${\overline{\rm MS}}$. 
 We show the ${s}_{n}^{NNA}$ coefficients in the 
${\overline{\rm MS}}$ and $V$-schemes up to $n=10$. 
 Further note that for $n>5$ the partial sums in the
two schemes are in complete agreement to the number of significant 
figures quoted, emphasising the
scheme-invariance of the resummed expansions in (138) and (139). 
In Table 7 the
partial sums for ${\alpha}_{s}^{\overline{MS}}=0.3$, 
(${\alpha}_{s}^{V}=0.46736$) appropriate
for the strange quark mass determination are recorded ($N_{f}=3$ is assumed). 
As one would
anticipate the convergence is much less impressive. 
Further the analytic continuation
terms are much more important at this larger value of the coupling, 
and the agreement of the
results in the two schemes, and the apparent convergence 
is much more evident for the
contour-improved expansion (139). 
This fact supports the application of the contour-improved NNLO expansions 
for the extraction of the $s$-quark mass value from the Cabibbo 
suppressed $\tau$-decay mode directly in the ${\overline{\rm MS}}$-scheme 
\cite{ChKuP} and within a realization of the 
effective charges approach \cite{r12}.
To emphasise the scheme-dependence of the all-orders sum of the 
``conventional'' NNA
expansions in (128) and (129) 
we tabulate the corresponding partial sums in Table 8, for
${\alpha}_{s}^{\overline{MS}}=0.114$. 
The partial sums in the two schemes are clearly
converging towards two different results, $0.04134$ in the 
$\overline{\rm MS}$ scheme and
$0.04039$ in the $V$-scheme. 
The difference ${\approx}0.001$ is of order ${({\alpha}_{s}/{\pi})}^{2}$, as
one would anticipate from (133).\\

We can compare the partial sums in Tables 6-8 with 
the all-orders Borel sum of the
series based on the Borel transform in (53). 
IR renormalons require regulation since they
contribute singularities on the positive real axis in the Borel plane. 
In common with
similar numerical studies on the vector correlator 
\cite{r7,r22,r24,Rid} we shall take a
Cauchy Principal Value ($PV$). In the $V$-scheme we then have
\begin{equation}
{\sum_{n>0}}{s}_{n}^{NNA,V}{\left(\frac{{\alpha}_{s}^{V}}{\pi}\right)}^{n}
{\equiv}PV
{\int_{0}^{\infty}}{d{\delta}}\;{e}^{-{\delta}{\pi}/({\beta}_{0}
{\alpha}_{s}^{V})}
\left[\frac{{\sin}{\pi}{\delta}}{{\pi}{\delta}}\;2\left
({G_+}({\delta})+{G_-}({\delta}) 
+\frac{{\gamma}_{0}}{\delta}\right)-\frac{2{\gamma}_{0}}{\delta}\right]\;,
\end{equation}
where the UV and IR renormalon contributions ${G_-}({\delta})$ and 
${G_+}({\delta})$ are
given by (44) and (45) respectively. 
Writing the `${\sin}$' as a sum of complex 
exponentials and using partial fractions the separate UV and IR 
renormalon contributions
can be expressed \cite{r22} in terms of generalized exponential 
integral functions
$Ei(n,w)$, defined for $Re w>0$ by 
\begin{equation}
Ei(n,w)={\int_{1}^{\infty}}{dt}\frac{{e}^{-wt}}{t^n}\;.
\end{equation}
One finds
\begin{eqnarray}
{\sum_{n>0}}{s}_{n}^{NNA,V}{\left(\frac{{\alpha}_{s}^{V}}{\pi}\right)}^{n}&=
&{A}_{0}^{NNA}({\alpha}_{s}^{V})-1
+4{A}_{1}^{NNA}({\alpha}_{s}^{V})
\nonumber \\
&-&\frac{4}{3{\pi}{\beta}_{0}}{\sum_{n>0}}\frac{{(-1)}^{n}}{n^2}
[{\phi_+}(1,n)+{\phi_+}(2,n)-{\phi_-}(1,n)-{\phi_-}(2,n)]
\nonumber \\
&+&\frac{1}{{\pi}{\beta}_{0}}\left[\frac{16}{3}{\phi_-}(1,1)-
\frac{7}{3}{\phi_-}(1,2)
+\frac{4}{3}{\phi_-}(2,1)-\frac{4}{3}{\phi_-}(2,2)\right]\;,
\end{eqnarray}
where ${\phi_+}$ and ${\phi_-}$ are defined by \cite{r22}
\begin{eqnarray}
{\phi_+}(p,q)&=&{e}^{q{\pi}/({\beta}_{0}{\alpha}_{s}^{V})}{(-1)}^{q}\;Im
\left[Ei\left(p,\frac{q{\pi}}
{{\beta}_{0}{\alpha}_{s}^{V}}+i{\pi}\right)\right]
\nonumber \\
{\phi_-}(p,q)&=&{e}^{-q{\pi}/({\beta}_{0}{\alpha}_{s}^{V})}{(-1)}^{q}\;
Im\left[Ei\left(p,-\frac
{q{\pi}}{{\beta}_{0}{\alpha}_{s}^{V}}-i{\pi}\right)\right]
\nonumber \\
&-&\frac{{e}^{-q{\pi}/({\beta}_{0}{\alpha}_{s}^{V})}{(-1)}^{q}{\pi}}{{(p-1)}!}
{\left(\frac{q}{{\beta}_{0}}\right)}^{p-1}\;
Re\left[{\left(\frac{\pi}{{\alpha}_{s}^{V}}+i{\pi}{\beta}_{0}
\right)}^{p-1}\right]\;.
\end{eqnarray}
Including 1000 terms in the sum over ${\phi_+}$ and ${\phi_-}$ in (142) 
gives a result
accurate to five significant figures. 
Exponentiating and evaluating $R$ from (138) then
yields the values indicated ``$PV$'' in the last row of Tables 6 and 7. 
The all-orders Borel sum in (140) is scheme-dependent by itself, but on
exponentiating it and combining with the factor of 
${({\alpha}_{s}/{\pi})}^{2{\gamma}_{0}/{\beta}_{0}}$ in
(138) one obtains a scheme-independent result by construction. 
In the $\overline{\rm MS}$ scheme
the Borel transform in (140) has an extra factor ${e}^{5{\delta}/3}$. 
At the smaller value
of the coupling ${\alpha}_{s}^{\overline{\rm MS}}=0.114$ in Table 6 one 
sees that
for $n>4$ the partial sums are all in good agreement with the  
exponentiated $PV$ Borel result.
The partial sums are in fact stable to four significant figures up to
$n{\approx}40$ where violent oscillations due to the leading 
UV renormalon
will be evident. The situation is somewhat less stable at the larger value
of ${\alpha}_{s}^{\overline{\rm MS}}=0.3$ in Table 7, with oscillations due to
the leading UV result clearly visible in the $V$-scheme. For
$n>4$ the $\overline{\rm MS}$ result is in reasonable agreement 
with the exponentiated
PV Borel sum, oscillations become evident for $n>9$, the $V$-scheme results 
break down
for $n>7$. The contour-improved  expansion results are stable but somewhat 
smaller than those
obtained with the standard perturbative expansion. On the evidence of 
these numerical comparisons
one would anticipate that, 
even at the larger value
of the coupling, fixed-order perturbation theory
at NNLO ($n=3$), the level of exact calculation at present, would give
a reasonable approximation. Finally, in Table 8 we give in the last 
row the $PV$ Borel
sums in the $\overline{\rm MS}$ and $V$ schemes. We see that these values are
in good agreement with the respective scheme-dependent 
``conventional NNA'' partial
sums for $n>4$.

\subsection{Effective charges from the scalar correlator}

In Sec.3.1 we discussed the problem of estimates 
of the higher-order QCD corrections to the spectral function 
of QCD sum rules in the scalar channel and to the 
decay width of a Standard Electroweak Model Higgs boson to 
quark-antiquark pairs within the NNA approach. In fact this problem
was already analysed a few years ago \cite{CKS} using a variant of the  procedure
developed in Refs.\cite{KS,KS2}, which is based on 
application of the effective charges approach of Ref.\cite{Grunberg1}. 
Usually this approach is applied to the renormalization scheme-dependent 
expansions of the quantities, which satisfy the renormalization group 
equations without anomalous dimension terms (see e.g. Eq.(66)).
However, the quantities of Eqs.(67),(71),(91) we are interested in 
obey the renormalization group equations with anomalous mass dimension 
function (see e.g. Eq.(49)), which arises due to the factor of two powers of the running quark mass
 appearing in the Born approximation of their expansions.
The appearance of the  anomalous dimension function in the corresponding 
renormalization group equations reflects the scale-scheme dependence 
of the running quark mass and it generates the additional scheme-dependence 
of the perturbative series under investigation (in contrast to 
the familiar $e^+e^-$ annihilation R-ratio, the scheme-dependence 
of $R_S$ is  starts from the ${\alpha_s}/\pi$-term). This additional 
scheme-dependence is analogous to the factorization scale-scheme dependence 
of the moments of deep-inelastic scattering structure functions (for some 
related discussions see Sec.3.4). In the process of the effective-charges 
motivated studies of Ref.\cite{CKS} the careful treatment of this 
``factorization-like'' scale-scheme ambiguity of the definition of running 
mass was overlooked. 
To overcome this shortcoming one should define the related effective charges 
either  using the representation of Eq.(123) of Sec.3.3 
\cite{Grunberg1,r18}  or defining the logarithmic derivative of the 
quantities under consideration \cite{Grunberg1}.
   
In this section we shall follow the latter prescription and consider  
the euclidean construct $R_D$ in~(\ref{RD}) which is the unique combination
of first and second derivatives of the high-energy scalar correlator that
satisfies a renormalization-group equation of the form~(\ref{RGD})
and gives $R_D=\alpha_s/\pi+O(\alpha_s^2)$.
Thus it provides the simplest way of defining an effective
coupling
\begin{equation}
{\widetilde{\alpha}_s(Q^2)\over\pi}\equiv R_D(Q^2)
\equiv-{1\over2}{d\log\widetilde{D}_S(Q^2)\over d\log Q^2}
=\sum_{n\ge0}
r_n\left({\alpha_s(Q^2)\over\pi}\right)^{n+1}
\label{eff}
\end{equation}
in the scalar channel. This leads to an effective beta function
\begin{equation}
\widetilde{\beta}(\widetilde{\alpha}_s(Q))
\equiv{d\log R_D(Q^2)\over d\log Q^2}=
-\sum_{n\ge0}\widetilde{\beta}_n
\left({\widetilde{\alpha}_s(Q^2)\over\pi}\right)^{n+1}
\label{beff}
\end{equation}
whose coefficients $\widetilde{\beta}_n$
are scheme-independent combinations of
the coefficients $\beta_n$ of the $\overline{\rm MS}$ beta function and the
coefficients $r_n$ of the $\overline{\rm MS}$ expansion of $R_D$.
Clearly $\widetilde{\beta}_n$ may differ from $\beta_n$ only for $n>1$.
To simplify the presentation, we define $c_n\equiv\beta_n/\beta_0$.
Then for $n>1$ we have
\begin{equation}
\frac{\tilde{\beta}_n-\beta_n}{n-1}=\beta_0(r_n-\Omega_n)
\label{om}
\end{equation}
where $\Omega_n$ is determined by products of elements of
$\{r_k,c_k\mid k<n\}$ with weights summing to $n$. In particular
\begin{eqnarray}
\Omega_2&=&r_1(c_1+r_1)\label{om2}\\
\Omega_3&=&r_1\bigg(c_2-\frac{1}{2}c_1r_1-2r_1^2+3r_2\bigg)\label{om3}\\
\Omega_4&=&r_1\bigg(c_3-\frac{4}{3}c_2r_1+\frac{2}{3}c_1r_2+
\frac{14}{3}r_1^3-\frac{28}{3}r_1r_2+4r_3\bigg)\nonumber\\&&{}
+\frac{1}{3}\bigg(c_2r_2-c_1r_3+5r_2^2\bigg)
\label{om4}
\quad{}\end{eqnarray}
to 5 loops (see \cite{KS,KS2}).

The estimation method based on this effective charge is
$r_4\approx\Omega_4$. It was 
motivated by the assumption that in electron-positron
annihilation and deep-inelastic scattering sum rules 
such scheme-independent process-dependent effective $\beta$ functions
will behave in a way that is broadly similar to the $\overline{\rm MS}$ 
$\beta$ function,
with $\widetilde{\beta}_n$ having the same sign
and magnitude as $\beta_n$ at 3 and 4 loops.
The calculation of Ref.\cite{4LB} confirmed this assumption
(for more detailed comparison of the behaviour of the effective charges 
$\beta$-functions for these processes with the $\overline{\rm MS}$
$\beta$-function at the 4-loop level see Ref.\cite{GK}).

Note however that the effective coupling $R_D(Q^2)$ contains the spurious
IR renormalon at $\de=1$, because it was constructed from the first
and second derivatives of the correlator, with the first requiring
UV subtraction for finite $m^2/Q^2$. The only way that
massless perturbation theory can tell us that we did something that would
be illegal in the massive theory is to make the massless perturbation
explode factorially, leaving an intrinsic ambiguity of order
$\Lambda^2/Q^2$ at the point where the sign-constant asymptotic series
becomes senseless. We emphasize that this $\Lambda^2/Q^2$
effect is profoundly perturbative and should not be confused
with suggestions of $\Lambda^2/Q^2$ effects beyond those
expected from the OPE \cite{Q2}. In our present case we know precisely why
$\widetilde{D}_S(Q^2)$ is sick: the affliction results from
perturbative malpractice, in failing to remove UV infinities
at finite mass. If one follows the good practice of~\cite{Becchi},
by taking the second derivative, then
the spurious $\de=1$ renormalon disappears.

It is therefore instructive to compare the performance of the $\Omega$
estimator with its obvious alternative
$\overline{r}_4\approx\overline{\Omega}_4$
based on the effective charge
\begin{equation}
\overline{R}_D(Q^2)
\equiv-{1\over2}{d\log\overline{D}_S(Q^2)\over d\log Q^2}
=\sum_{n\ge0}
\overline{r}_n\left({\alpha_s(Q^2)\over\pi}\right)^{n+1}
\label{effb}
\end{equation}
where $\overline{r}_n$ is obtained by replacing $d_n$
in~(\ref{r0})-(\ref{r4}) by $\overline{d}_n$ from~(\ref{dbn}).
This effective charge is the unique choice formed from the
second and third derivatives of the correlator and is hence the
simplest that is free from the spurious renormalon.
To compute $\overline{\Omega}_n$ exactly
one merely replaces $r_n$ by $\overline{r}_n$ in~(\ref{om2})--(\ref{om4}).

In Tables 9 and 10 we compare the estimators $\Omega$ and $\overline{\Omega}$
with exact results at 3 and 4 loops. In Table~9 we also give
the naive nonabelianizations of $r_n$ and $\Omega_n$.
Those for $\overline{r}_n$ and $\overline{\Omega}_n$
are in Table~10. Here
\begin{eqnarray}
r_n^{\rm NNA}&=&{n\over2}\beta_0^n\widetilde{\Delta}_{n+1}
\label{rnna}\\
\overline{r}_n^{\rm NNA}&=&{n\over2}\beta_0^n\overline{\Delta}_{n+1}
\label{rbnna}
\end{eqnarray}
are obtained by $N_f\to N_f-33/2$ in~(\ref{rl})
and its corresponding version for the effective charge $\overline{R}_D(Q^2)$.
The NNA expressions for the estimators $\Omega$ are 
\begin{eqnarray}
\Omega_2^{NNA} &=& \bigg(r_1^{NNA}\bigg)^2 \\ \nonumber 
\Omega_4^{NNA} &=& 3r_2^{NNA}r_1^{NNA}-2\bigg(r_1^{NNA}\bigg)^3 \\ \nonumber 
\Omega_4^{NNA} &=& 4r_3^{NNA}r_1^{NNA}
-\frac{28}{3}r_2^{NNA}\bigg(r_1^{NNA}\bigg)^4
+\frac{14}{3}\bigg(r_1^{NNA}\bigg)^4+\frac{5}{3}\bigg(r_2^{NNA}\bigg)^2
\end{eqnarray}
or 
\begin{eqnarray}
\Omega_2^{NNA}&=&\frac{1}{4}\beta_0^2\widetilde{\Delta}_2^2 \\ \nonumber
\Omega_3^{NNA}&=&\frac{1}{4}\beta_0^3\bigg(6\widetilde{\Delta}_3\widetilde{\Delta}_2
-\widetilde{\Delta}_2^3\bigg) \\ \nonumber 
\Omega_4^{NNA}&=&\beta_0^4\bigg(3\widetilde{\Delta}_4\widetilde{\Delta}_2
-\frac{7}{3}\widetilde{\Delta}_3\widetilde{\Delta}_2^2
+\frac{7}{24}\widetilde{\Delta}_2^4+\frac{5}{3}\widetilde{\Delta}_3^2\bigg)
\label{omega}
\end{eqnarray}
while the ones for $\overline{\Omega}$ can be obtained by 
$r_i^{NNA}$ to $\overline{r}_i^{NNA}$ in Eqs.(153)  or 
$\widetilde{\Delta}_n$ to $\overline{\Delta}_n$ in Eqs.(154).
For example, the analytical form of the  NNA expressions for 
estimators $\Omega_4$ and $\overline{\Omega}_4$ read
\begin{eqnarray}
\Omega_4^{\rm NNA}&=&\beta_0^4\left(
{80\over3}[\zeta(3)]^2 + {44\over9}\zeta(3) + {124927\over324}\right)
\label{Omn4}\\
\overline{\Omega}_4^{\rm NNA}&=&\beta_0^4\left(
{80\over3}[\zeta(3)]^2 + {668\over9}\zeta(3)+ {33463\over324}\right)
\label{Ombn4}
\end{eqnarray}
They result from neglect of $c_k$
in~(\ref{om4}), since $c_k=O(N_f^{k-1})$, at large $N_f$,
while $r_k$ and $\overline{r}_k$ are $O(N_f^k)$.

In Table~9 we present results using the $\Omega$ estimator.
The comparison with exact results at 3 and 4 loops 
is not so good as in the QCD and QED studies of Refs.\cite{KS,KS2} correspondingly.
At $N_f=5$, for example, we have $r_2/\Omega_2\approx0.702$,
while $r_3/\Omega_3\approx2.34$, so the exact growth factor $r_3/r_2$
is $2.34/0.702\approx3.33$ times that estimated by $\Omega$.
Considering the NNA approximations  for $r_n$ and $\Omega_n$ we 
get the similar results.
Thus effective-charge analysis, 
based on both $\Omega$ and $\Omega^{NNA}$,
give the results  over factor 3 larger than the growth factor estimated
by~(\ref{rnna}).

At $N_f=5$, this gives
$r^{\rm NNA}_2/r_2\approx0.60$
and $r^{\rm NNA}_3/r_3\approx0.59$,
with an actual growth factor $r_3/r_2$ only 2\% greater than that indicated
at large-$\beta_0$. This is in accord with the global pattern
of Table~4, which shows that NNA gives a reasonable account of growth
from 3 to 4 loops in all 3 quantities that were considered there.
Now we turn attention to the final 3 rows of Table~9.
The values of $\Omega_4$ in the final row are exact, since
by definition they entail only input from lower orders of perturbation theory.
The value of $r_4$ is quite unknown: the $\Omega$ method takes
$\Omega_4$ as its estimate. The smaller values of $r_4^{\rm NNA}$
are simply obtained by replacing $N_f$ by $N_f-33/2$ in the exact
large-$N_f$ result for $r_4$. The intermediate values
of $\Omega^{\rm NNA}_4$ come from~(\ref{Omn4}).
The $N_f$-independent ratio
$r^{\rm NNA}_4/\Omega^{\rm NNA}_4\approx0.46$
is a precise measure of the uncertainty 
of the $\Omega$ estimator
at large $N_f$.
The $N_f$-dependent ratio
$\Omega^{\rm NNA}_4/\Omega_4\approx0.5$ is a precise measure of the 
uncertainty 
of NNA for $N_f=3,4,5$.
These two effects lead to
a factor of 4 difference between the estimators $r^{\rm NNA}_4$
and $\Omega_4$, as rival candidates for $r_4$.

Now we turn to Table~10, where the $\overline{\Omega}$
method fails to get the sign right for $r_3$, at 4 loops.
Inspecting the
final 3 rows, we see that the $\overline{\Omega}$
estimator is $10^3$ times its target at large $N_f$.
On the other hand, the NNA estimate~(152) is very successful.
In view of the these failings of $\overline{\Omega}$,
we proceed only with the $\Omega$ estimator.

To convert a prediction
of $r_4$ into one for the 5-loop term $s_4$ in
the physically relevant imaginary part, we
use the known terms in the quartic
\begin{eqnarray}
r_4-\gamma_4-2\beta_0s_4&=&
37136.85285-7810.216455N_f+575.3282994N_f^2\nonumber\\&&{}
-16.32062026N_f^3+0.1444281007N_f^4
\label{oms4}
\end{eqnarray}
and parametrize errors in the $\Omega$ estimator by
$r_n/\Omega_n=1+\delta_n$.
Then we obtain
\begin{eqnarray}
s_4&=&\widetilde{s}_4-\gamma_4/2\beta_0\label{s4h}\\
\widetilde{s}_4(N_f=3)&=&\phantom{-}472+4574\delta_4\label{s4h3}\\
\widetilde{s}_4(N_f=4)&=&\phantom{-}146+3529\delta_4\label{s4h4}\\
\widetilde{s}_4(N_f=5)&=&         - 129+2615\delta_4\label{s4h5}
\end{eqnarray}
with central values that are small compared with
a realistic estimate of the uncertainty, bearing in mind that
$\delta_2\approx-0.3$ and $\delta_3\approx1.3$, at $N_f=5$.

As an indication of the problem at 3 and 4 loops,
we give the 4-loop effective beta
function in terms of $\tilde{a}_s\equiv R_D$ at $N_f=3,4,5$:
\begin{eqnarray}
\widetilde{\beta}(N_f=3)&=&-2.250\tilde{a}_s
-4.000\tilde{a}_s^2+58.920\tilde{a}_s^3-2148.503\tilde{a}_s^4
\label{beff3}\\
\widetilde{\beta}(N_f=4)&=&-2.083\tilde{a}_s
-3.208\tilde{a}_s^2+53.852\tilde{a}_s^3-1687.191\tilde{a}_s^4
\label{beff4}\\
\widetilde{\beta}(N_f=5)&=&-1.917\tilde{a}_s
-2.417\tilde{a}_s^2+49.356\tilde{a}_s^3-1303.490\tilde{a}_s^4
\label{beff5}
\end{eqnarray}
with a sign change at 3 loops as a result of the appearance of a 
large and positive 3-loop coefficient. This pattern was 
already observed to occur in the effective beta-function for the minkowskian 
analog 
of $R_D$ in Ref.\cite{Gor3}, where it was considered as an indication 
of the existence of the spurious perturbative infrared fixed point.
Indeed, this zero  is compensated by the 4-loop
terms, which remove the spurious fixed point (for the demonstration
of the existence of a similar feature in the minkowskian region see 
Ref.\cite{4LM2}).

They are however $O(50)$ times larger than those in the $\overline{\rm MS}$-scheme 
beta functions
\begin{eqnarray}
\beta(N_f=3)&=&-2.250a_s-4.000a_s^2          -10.060a_s^3-47.228a_s^4
\label{bet3}\\
\beta(N_f=4)&=&-2.083a_s-3.208a_s^2-\phantom{1}6.349a_s^3-31.387a_s^4
\label{bet4}\\
\beta(N_f=5)&=&-1.917a_s-2.417a_s^3-\phantom{1}2.827a_s^3-18.852a_s^4
\label{bet5}
\end{eqnarray}
with $a_s\equiv\alpha_s/\pi$.

From this perspective, it is unsurprising that the $\Omega$
estimator 
performs less reliably 
at 3 and 4 loops than in the 
cases considered in Refs.\cite{KS,KS2}: at 3 loops it is bound to
overestimate $r_2$, because of the sign change of $\widetilde{\beta}_2$; at 4 loops
it is bound to underestimate $r_3$, because
$\widetilde{\beta}_3/\beta_3^{\overline{\rm MS}}\approx50$, 
whereas the procedure assumes that
it is $O(1)$.

We conclude that the application of $\Omega$ estimator 
in the scalar channel allows values of the 5-loop
coefficient $s_4$ between $-10^3$ and $+10^3$, if the accuracy is
no better than $\delta_4=0\pm0.3$, i.e.\ no better than
at 3 loops. If it is no better than at 4 loops, the range
widens further, by a factor of about 3.

The appearance of large and negative estimates for $s_4$ in Ref.\cite{CKS},
which contradict the results of application of the NNA procedure, described 
in detail in Sec.3.1, is a reflection of the similar problem 
encountered in Ref.\cite{CKS} in a simplified variant of the effective-charges 
approach in the same scalar channel.

\section{Conclusions}
In this paper we have extended the existing large-$N_f$ analysis of
the vector correlator [14] to the previously uninvestigated case of
the scalar correlator. Because of the absence of the Ward identity
${Z_1}={Z_2}$ present in the vector case, and the inevitable involvement
of the quark mass anomalous dimension, the analysis amd combinatorics
was considerably more complicated. An all-orders large-$N_f$ result
for the anomalous mass dimension ${\gamma}_{m}$ was obtained in (39,40),
and thanks to the remarkable identity (20) relating insertions into
two-loop skeleton diagrams to the anomalous dimension, it was possible
to obtain the all-orders large-$N_f$ result for the coefficient 
function of $R_S$ in (51)-(53). As in the vector case the Borel
transform of (44) and (45) contained  UV and IR renormalons. A new feature
of the scalar analysis was the presence of a leading IR renormalon
at ${\delta}=1$ in (45). This is not present in the physical quantity
$R_S$ thanks to the analytical continuation factor ${\sin}({\pi}{\delta})/
{\pi}{\delta}$ in (53), and the leading singularity is the UV renormalon
at ${\delta}=-1$.
 It is, however, present in the quantity
$\widetilde{D}_{S}({Q}^{2})$ of (64), which would correspond to the
obvious generalization of the vector Adler function , related by
a singly subtracted dispersion relation to the scalar vacuum polarization
${\Pi}_{S}$. The presence of this leading IR renormalon is connected
with the fact that the undetermined constant 
${\Pi}_{S}(0)$ 
in ${\Pi}_{S}$ is
{\it infinite}, a circumstance which did not occur in the vector
case thanks to the ${Z_1}={Z_2}$ Ward identity. A more satisfactory
choice for the scalar Adler function is therefore the twice subtracted
construct ${\overline{D}}_{S}$ defined in (63). An initial survey of
the growth of the coefficients $S_n$ in (50) was  given in Table 2, where
it is seen that in the ${\overline{\rm MS}}$ scheme for $n<7$ there is
rather stable behaviour, with rapid growth corresponding to the leading
UV renormalon evident for $n>7$. The perturbation series for
${\overline{D}}_{S}$,${\widetilde{D}}_{S}$ and its logarithmic derivative
$R_D$ are studied in Sec. 2.10. In Sec.3 we moved  to a study of
so-called ``naive non-abelianization'' (NNA) , where $N_f$ in the
large-$N_f$ result is replaced by ${N_f}-11{N_c}/2=-6{\beta}_{0}$, and
the pieces of perturbative coefficients  containing the leading
power of ${\beta}_{0}$ are resummed to all-orders. For the quantities
${\overline{D}}_{S}$,${\widetilde{D}}_{S}$ and $R_S$ the use of NNA to
estimate the known three and four-loop coefficients was found to give
the correct sign and order of magnitude. 
Moreover, the NNA-inspired estimates predict that at the five-loop level 
the numerical values of the corresponding coefficients are positive and 
not very large. In the case of the imaginary part of the scalar correlator, 
related to the Higgs boson decay width to quark-antiquark pairs, 
and to the QCD Sum Rules spectral functions, we have 
$s_4^{NNA}(N_f=3)\approx 49$, $s_4^{NNA}(N_f=4)\approx 39$, 
$s_4^{NNA}(N_f=5)\approx 31$. Using the conservative 
estimate of an overall factor of two for the uncertainty of the
NNA estimates in the scalar channel (suggested by careful comparison
of the three and four-loop NNA estimates to the results of
explicit calculation, see Table 4) 
we conclude that the NNA-inspired  estimate 
$s_4\approx 2s_4^{NNA}$ gives us the following numbers $s_4(N_f=3)\sim 98$, 
$s_4(N_f=4)\sim 78$, $s_4(N_f=5)\sim 62$.
For $N_f=5$ our estimate is in good agreement with  
the result of previous studies, based on application 
of the  [2/2] asymptotic Pad\'e estimation technique [24].
However, for $N_f=3$ the method of Ref.\cite{HPade}
agrees with our estimates only in sign and is 
over 2.6 times larger than the estimates proposed 
by us.
In Sec. 3.2 we noted that for the vector
correlator the NNA term does have some special properties, deriving
from an analysis of the operators that build the leading UV 
renormalon singularity. On re-expansion it reproduces the
sub-leading in $N_f$ contributions to asymptotic accuracy
O$(1/n)$ in ${n}^{\rm{th}}$ order perturbation theory. A similar
result holds for a ``dual-NNA'' , exact in the large-$N_c$ limit.
For the scalar case a corresponding weak asymptotic result was not
expected to hold, but the ``dual NNA'' term should still provide
a good approximation. In Sec. 3.3 we considered the analytical
continuation from ${\widetilde{D}}_{S}$ to $R_S$ . We recast the
running mass $m({Q}^{2})$ in terms of the RG-invariant mass
${\hat{m}}$ . The analytical continuation was similar to
the much studied Euclidean to Minkowskian continuation for the
${e}^{+}{e}^{-}$ $R$-ratio [29-40]. The presence of an anomalous 
dimension complicated the analysis slightly. We arrived at the
``contour-improved'' expansion for $R_S$ in (129), in which
a subset of analytical continuation terms involving ${\pi}^{2}$ are
resummed to all-orders, at each order of the expansion. A further
subtlety due to the presence of an ${{\alpha}_{s}}^{2{\gamma}_{0}/
{\beta}_{0}}$ term involving the anomalous dimension, was  that the
straightforward NNA expansions in (128) and (129) have all-orders
sums which are RS-dependent. We showed how this could be
remedied in Sec. 3.4 , and obtained the reformulated expansions
in (138) and (139) whose all-orders sums were scheme-independent.
In 3.5 we performed comparisons of fixed-order perturbation theory 
with the all-orders sum defined using a Cauchy principal value
of the Borel sum, based on the Borel transform of (44),(45),(Tables
6-8). Two values of the coupling, ${\alpha}_{s}^{\overline{\rm MS}}=0.114$
and ${\alpha}_{s}^{\overline{MS}}=0.3$, appropriate for
the calculation of the Higgs decay width to a quark-antiquark pair,
and for the strange quark mass determination from QCD Sum Rules,
respectively, were considered. The ${\overline{\rm MS}}$ scheme and
$V$-scheme were  used to illustrate the scheme-dependence issue. Even at
the larger value of the coupling it seemed that satisfactory accuracy
was achieved at order $n=3$, the highest  order at which exact calculations
are so far available.   
However, from the analysis of Ref.\cite{Yndc} one can conclude that 
the uncertainties of the existing $m_s$ extractions from the 
QCD sum rules for the scalar correlator might be underestimated.
In view of our analysis we conclude, that these possible additional 
uncertainties  are not coming from the  uncalculated 
5-loop perturbative QCD contributions, but are mainly related to
the uncertainty  of the experimental model for the spectral 
function of the QCD sum rules   in the scalar 
channel.\\

In overall conclusion we have provided a framework for extending 
the numerous published investigations of higher-order perturbative
behaviour, analytical continuation, approximate all-orders
resummation, and estimates of higher-order corrections, 
for the vector correlator and its derived quantities,
to the much less studied case of the scalar correlator. We hope
that our results may be of use in further phenomenological
investigations.

\section {Note Added in Proof}

After this work was submitted for publication we were informed that 
the 3-loop corrections to the correlator of scalar currents have been 
calculated up to $(m^2/Q^2)^4$-terms \cite{95}. The application of 
Pad\'e resummation technique \cite{96,97} allowed the authors 
of Ref.\cite{98} to specify the excat mass-dependence of the scalar-scalar
correlator at 3-loops in a semi-analytical way. The analytical mass 
dependence of the 3-loop double-bubble contribution to the scalar 
correlator was studied in Ref.\cite{99}. It should be stressed, however, 
that none of these calculations affect the results reported in this our work.

Another, more closely related investigation, was performed in Ref.\cite{100},
where the effects of the UV-renormalons to the IR-safe Adler function 
of Eq.(63) of the scalar correlator were studied both in the 1-renormalon 
and 2-renormalon chain approximations. However, this analysis was based not 
on the exact calculations of the renormalon chain diagrams in the large 
$N_f$-limit, but on the latge $N_c$-analysis of the contributing 4-fermion 
operators. Moreover, in Ref.\cite{100} the analysis of the IR renormalon
contributions was not considered. In view of these differencies, 
it would be of interest to compare our results with those 
of Ref.\cite{100} in more detail.

\section{Acknowledgements}
We would like to thank A. Mirjalili for crucial help in producing and
checking the results in Tables 5-8, and M. Beneke for some 
clarifying remarks.

The preliminary part of this work was reported in the talks 
of two of us (DJB and ALK) at Quarks-2000 International Seminar, 
Pushkin, May 2000.
DJB is grateful to the organizers of this Seminar for hospitality 
in Pushkin and St.Petersburg.

The authors are  grateful to the participants 
of Quarks-2000 , and in particular,  A.A. Pivovarov, 
A.N. Tavkhelidze, L.G. Yaffe and F.J. Yndurain  for 
their interest to the work and useful 
discussions. We also would like to acknowledge the informal and constructive 
comments of L.D. Soloviev.

It is also a pleasure to thank E. de Rafael and H. K\"uhn for informing us 
of some additional works related to the content of our paper, after 
its submission for publication.

The essential part of this work was done when one of us (ALK) was 
visiting the Centre for Particle Theory of the University of Durham.
He would like to thank the UK Royal Society for financial support 
and his colleagues from Durham for hospitality.
While in Russia the work of ALK was done within the 
framework of scientific projects of  RFBR Grants N 99-01-00091, 
N 00-02-17432.

\newpage

\begin{center}{\bf Table~1:} Renormalons in~(\ref{Hr})\end{center}
$$\begin{array}{|r|rrr|rrr|r|}\hline n&
H^{\overline{\rm MS}}_n&H_n^{(0)}&H_n^{(-)}
&H_n^{(1)}&H_n^{(2)}&H_n^{(+)}&H_n\\\hline
2&19.625&-12.500&     2.594&    -48.000&    9.750&   1.043&     -27.488\\
3&11.093&  6.944&     5.870&     73.000&  -11.125&  -1.306&      84.476\\
4&-8.684& -5.787&    12.276&   -173.778&   18.069&   2.289&    -155.615\\
5&-0.056&  5.787&    45.579&    556.056&  -35.785&  -4.764&     566.817\\
6& 1.882& -6.430&   207.742&  -2270.519&   83.144&  11.212&   -1972.969\\
7&-0.509&  7.655&  1177.991&  11416.893& -223.934& -29.266&   12348.830\\
8&-0.142& -9.569&  7883.559& -68593.216&  694.766&  83.893&  -59940.709\\
9& 0.095& 12.404& 60911.356& 480286.469&-2465.171&-262.531&  538482.621\\
\hline\end{array}$$

\vfill

\begin{center}{\bf Table~2:} Renormalons in~(\ref{Sn})\end{center}
$$\begin{array}{|r|rrr|rrr|r|}\hline n&
S^{\overline{\rm MS}}_n&S_n^{(0)}&S_n^{(-)}
&S_n^{(1)}&S_n^{(2)}&S_n^{(+)}&S_n\\\hline
 2&-1.667&     3.333&     1.097&      4.000&   -1.000&    -0.097&      5.667\\
 3& 0.972&    -0.512&    -0.576&     10.667&   -2.167&    -0.232&      8.152\\
 4& 0.834&    -7.880&    -0.999&     19.285&   -1.655&    -0.263&      9.323\\
 5& 0.055&   -13.817&    -2.495&     10.576&    9.338&     0.762&      4.419\\
 6&-0.155&     9.160&    10.385&    -68.229&   46.309&     5.344&      2.813\\
\hline
 7&-0.056&   105.021&   -17.734&   -249.544&   92.873&    15.175&    -54.266\\
\hline
 8& 0.006&   200.490&   183.066&   -237.016&  -36.445&    13.543&    123.645\\
 9& 0.008&  -255.017& -1342.424&   1147.753& -829.274&   -87.712&  -1366.667\\
10& 0.001& -2257.800&  9978.990&   5101.747&-2297.027&  -516.006&  10009.905\\
11& 0.000& -4046.531&-91303.959&   5275.323& -176.520& -1500.459& -91752.148\\
\hline\end{array}$$

\vfill

\begin{center}{\bf Table~3:} Contributions to~(\ref{dtl},\ref{sl},\ref{dbl})
\end{center}$$\begin{array}{|l|r|rrr|}\hline n&A_n\quad{}&
\widetilde{\Delta}_n\quad{}&\Delta_n\quad{}&\overline{\Delta}_n\quad{}\\\hline
2 & - 1.66667 & 7.33333 &  7.33333 &  5.33333\\
3 &   0.97222 & 10.4696 &  7.17968 &  3.13622\\
4 &   0.83422 & 32.6145 &  8.48885 &  11.6754\\
5 &   0.05531 & 97.9534 &  4.36402 &  0.10978\\
6 & - 0.15502 & 503.887 &  2.96849 &  112.074\\
7 & - 0.05583 & 2194.28 & -54.2101 & -325.157\\
8 &   0.00581 & 16465.8 &  123.639 &  3300.11\\
\hline\end{array}$$

\newpage

\begin{center}{\bf Table~4:}
Ratios of NNA estimates to exact results
\end{center}
$$\begin{array}{|l|r|r|r|}
 \hline
     &                   N_f=3 &    N_f=4 &    N_f=5\\
 \hline
 d^{\rm NNA}_2/d_2                       & 0.514 & 0.496 & 0.477\\
 s^{\rm NNA}_2/s_2                       & 0.507 & 0.490 & 0.472\\
 \overline{d}^{\rm NNA}_2/\overline{d}_2 & 0.498 & 0.484 & 0.469\\
 \hline
 d^{\rm NNA}_3/d_3                       & 0.354 & 0.346 & 0.339\\
 s^{\rm NNA}_3/s_3                       & 0.482 & 0.565 & 0.747\\
 \overline{d}^{\rm NNA}_3/\overline{d}_3 & 0.764 & 0.871 & 1.081\\
 \hline
\end{array}$$

\vfill

\begin{center}{\bf Table~5:} ${s}_{n}^{NNA}$ coefficients in 
the $\overline{\rm MS}$ and
$V$-schemes\end{center} 
$$\begin{array}{|r|r|r|}\hline n&\overline{\rm MS}&V\\\hline
1&7.3333&4\\
2&13.761&-4.3408\\
3&31.184&6.7614\\
4&30.727&-58.738\\
5&40.061&456.674\\
6&-1402.21&-4161.77\\
7&6129.65&47635.6\\
8&-129864.7&-640093.5\\
9&1.8231 \;{10}^{6}&9.8022 \;{10}^{6}\\
10&-3.2028\; {10}^{7}&-1.6900 \; {10}^{8}\\
\hline \end{array}$$

\vfill

\begin{center}{\bf Table~6:} Partial sums of (137,138) for 
${\alpha}_{s}^{\overline{\rm MS}}=0.114$ 
\end{center}
$$\begin{array}{|r|rr|rr|}\hline n&
{R}^{(n)}_{\overline{\rm MS}}&{R}^{(n)}_{V}&
{R}^{(n)CI}_{\overline{\rm MS}}&{R}^{(n)CI}_{V}\\\hline
1&0.04099&0.04210&0.04049&0.04153\\
2&0.04174&0.04179&0.04152&0.04166\\
3&0.04180&0.04181&0.04180&0.04174\\
4&0.04181&0.04180&0.04180&0.04179\\
5&0.04181&0.04181&0.04180&0.04181\\
6&0.04181&0.04181&0.04180&0.04180\\
7&0.04181&0.04181&0.04181&0.04181\\
8&0.04181&0.04181&0.04181&0.04181\\
9&0.04181&0.04181&0.04181&0.04181\\
10&0.04181&0.04181&0.04181&0.04181\\
.&.......&.......&.......&.......\\
PV&0.04179&0.04179&0.04179&0.04179\\

\hline \end{array} $$

\newpage

\begin{center}{\bf Table~7:} Partial sums of (137,138) for 
${\alpha}_{s}^{\overline{\rm MS}}=0.3$ 
\end{center}
$$\begin{array}{|r|rr|rr|}\hline n&
{R}^{(n)}_{\overline{\rm MS}}&R^{(n)}_{V}
&{R}^{(n)CI}_{\overline{\rm MS}}&{R}^{(n)CI}_{V}\\\hline
1&0.24971&0.33333&0.21619&0.25609\\
2&0.28934&0.29778&0.25056&0.26237\\
3&0.30037&0.30706&0.26816&0.27759\\
4&0.30162&0.29310&0.27509&0.27676\\
5&0.30180&0.31222&0.27853&0.27606\\
6&0.30109&0.28234&0.27906&0.27800\\
7&0.30144&0.34521&0.27765&0.26881\\
8&0.30061&0.21538&0.27433&0.27246\\
9&0.30191&0.76056&0.26774&0.27747\\
10&0.29934&0.01704&0.26192&0.25527\\
.&.......&.......&.......&.......\\
PV&0.30138&0.30138&0.30138&0.30138\\
\hline \end{array} $$

\vfill

\begin{center}{\bf Table~8:} Partial sums of the ``conventional'' 
NNA (128,129) for ${\alpha}_{s}^{\overline{\rm MS}}
=0.114$
\end{center}
$$\begin{array}{|r|rr|rr|}\hline n&
{R}^{(n)}_{\overline{\rm MS}}&{R}^{(n)}_{V}&{R}^
{(n)CI}_{\overline{\rm MS}}&{R}^{(n)CI}_{V}\\\hline
1&0.03977&0.04159&0.03939&0.04110\\
2&0.04034&0.04133&0.04018&0.04121\\
3&0.04039&0.04135&0.04034&0.04133\\
4&0.04039&0.04134&0.04038&0.04133\\
5&0.04039&0.04134&0.04039&0.04134\\
6&0.04039&0.04134&0.04039&0.04134\\
7&0.04039&0.04134&0.04039&0.04134\\
8&0.04039&0.04134&0.04039&0.04134\\
9&0.04039&0.04134&0.04039&0.04134\\
10&0.04039&0.04134&0.04039&0.04134\\
.&.......&.......&.......&.......\\
PV&0.04038&0.04133&0.04038&0.04133\\
\hline \end{array} $$

\newpage

\begin{center}{\bf Table~9:}
Exact values of $r_n$ and $\Omega_n$,
with NNA estimates of $r_n$ and $\Omega_n$
\end{center}
$$\begin{array}{|l|r|r|r|}
 \hline
     &                   N_f=3 &    N_f=4 &    N_f=5\\
 \hline
 r_1 &                   10.17 &     9.56 &     8.94\\
 r_1^{\rm NNA}          &             8.25  &     7.64 &     7.03 \\
\hline
 r_2 &                   90.78 &    77.13 &    64.06\\

 r_2^{\rm NNA}  &         53.00    &     45.44 &   38.46 \\ 

\Omega_2 &             121.44 &   106.02 &    91.28\\
 
\Omega_2^{\rm NNA}  &     68.06 & 58.35 & 49.39 \\ 

\hline
 r_3 &                 1091.55 &   822.19 &   585.53\\
 
 r_3^{\rm NNA} & 557.25 & 442.36 & 344.46 \\ 

\Omega_3 &             620.75 &   424.80 &   250.41\\

\Omega_3^{\rm NNA}  & 188.77 & 149.85 & 116.69 \\

 \hline
 r^{\rm NNA}_4 &       5020.88 &  3690.50 &  2643.84\\
 \Omega^{\rm NNA}_4 & 11020.06 &  8100.07 &  5802.84\\
 \Omega_4 &           20583.50 & 14704.33 & 10024.21\\
 \hline
\end{array}$$

\vfill

\begin{center}{\bf Table~10:}
Exact values of $\overline{r}_n$ and $\overline{\Omega}_n$,
with NNA estimates of $\overline{r}_n$ and $\overline{\Omega}_n$
\end{center}
$$\begin{array}{|l|r|r|r|}
 \hline
     &                   N_f=3 &    N_f=4 &    N_f=5\\
 \hline
 \overline{r}_1 &                     7.92 &    7.47 &     7.03\\
 
\overline{r}_1^{\rm NNA}  & 6.00 & 5.56 & 5.11 \\

\hline
 \overline{r}_2 &                    36.53 &   29.94 &    23.52\\
 
\overline{r}_2^{\rm NNA}  & 15.88 & 13.61 & 11.52 \\

\overline{\Omega}_2 &               76.75 &   67.34 &    58.25\\
 
\overline{\Omega}_2^{\rm NNA}   &     36.00 & 30.86 & 26.12 \\

\hline
 \overline{r}_3 &                   233.13 &  138.29 &    55.79\\

\overline{r}_3^{\rm NNA}  &     199.49  & 158.36 & 123.31 \\

 \overline{\Omega}_3 &             -145.01 & -183.52 &  -219.13\\
 
\overline{\Omega}_3^{\rm NNA}  & -146.21 & -116.07 & -90.38\\

\hline
 \overline{r}^{\rm NNA}_4 &           5.63 &    4.14 &     2.96\\
 \overline{\Omega}^{\rm NNA}_4 &   5921.10 & 4352.18 &  3117.87\\
 \overline{\Omega}_4 &             6619.34 & 4648.70 &  3131.62\\
 \hline
\end{array}$$

\newpage
\raggedright


\begin{thebibliography}{99}

\bibitem{KC}  K.G. Chetyrkin, {\it Phys. Lett.} {\bf B390} (1997) 309.
\bibitem{GKLR} S.G. Gorishny, A.L. Kataev and S.A. Larin, 
{\it Phys. Lett.} {\bf B273} (1991) 141;\\
L.R. Surguladze and M.A. Samuel, {\it Phys. Rev. Lett.} {\bf 66} (1991) 560; 
ibid. {\bf 66} (1991) 2416 (Err);\\
K.G. Chetyrkin, {\it Phys. Lett.} {\bf B390} (1997) 309.
\bibitem{DJB} D.J. Broadhurst, {\it Phys. Lett.} {\bf B101} (1981) 423.
\bibitem{Gor1} S.G. Gorishny, A.L. Kataev and S.A. Larin, 
{\it Sov. J. Nucl. Phys.} {\bf 40} (1984) 329.
\bibitem{Gor2}
S.G. Gorishny, A.L.Kataev, S.A. Larin and L.R. Surguladze, 
{\it Mod. Phys. Lett.} {\bf A5} (1990) 2703.
\bibitem{Gor3} 
S.G. Gorishny, A.L. Kataev, S.A. Larin and L.R. Surguladze, 
{\it Phys. Rev.} {\bf D43} (1991) 1633.
\bibitem{3M2}
A.L. Kataev and V.T. Kim, Preprint ENSLAPP-A-407-92 (hep-ph/9304282);
{\it Mod. Phys. Lett.} {\bf A9} (1994) 1309;\\
L.R. Surguladze, {\it Phys. Lett.} {\bf B341} (1994) 60;\\
K.G. Chetyrkin and A. Kwiatkowski, {\it Nucl. Phys.} {\bf B461} (1996) 3;\\
S.A. Larin, T. van Ritbergen and J.A.M. Vermaseren, 
{\it Phys. Lett.} {\bf B362} (1995) 134.
\bibitem{ST}
L.R. Surguladze and F.V. Tkachov, {\it Nucl. Phys.} {\bf B331} (1990) 35.
\bibitem{CPS}  K.G. Chetyrkin, D. Pirjol and K. Schilcher, 
{\it Phys. Lett.} {\bf B404} (1997) 337. 
\bibitem{ChKSt}
K.G. Chetyrkin, B.A. Kniehl and M. Steinhauser, 
{\it Phys. Rev. Lett.} {\bf 79} (1997) 353.
\bibitem{4LM1} K.G. Chetyrkin, {\it Phys. Lett.} {\bf B404} (1997) 161.
\bibitem{4LM2} J.A.M. Vermaseren, S.A. Larin and T. van Ritbergen,
{\it Phys. Lett.} {\bf B405} (1997) 327.
\bibitem{4LB} T. van Ritbergen, J.A.M. Vermaseren and S.A. Larin, 
{\it Phys. Lett.} {\bf B400} (1997) 379. 
\bibitem{LNF} D.J. Broadhurst, {\it Z. Phys.} {\bf C58} (1993) 339.
\bibitem{BG}  D.J. Broadhurst and A.G. Grozin, {\it Phys. Rev.} {\bf D52} 
(1995) 4082.  
\bibitem{Zakh} V.I. Zakharov, {\it Nucl. Phys.} {\bf B385} (1992) 452.
\bibitem{BY} L.S. Brown and L.G. Yaffe, {\it Phys. Rev.} {\bf D45} (1992) R398;\\
L.S. Brown, L.G. Yaffe and Ch. Zhai, {\it Phys. Rev.} {\bf D46} (1992) 4712.
\bibitem{Chris} C.N. Lovett-Turner and C.J. Maxwell, {\it Nucl. Phys.} 
{\bf B432} (1994) 147.
\bibitem{Alt} G. Altarelli, ``Introduction to renormalons'', CERN-TH-95-309;
Proceedings of Int. School of Subnuclear Physics: 33rd Course: 
Vacuum and Vacua: The Physics of Nothing, 2-10 July 1995, Erice, Italy.
\bibitem{BenekeR} M. Beneke, {\it Phys. Rept.} {\bf 317} (1999) 1. 
\bibitem{KS} A.L. Kataev and V.V. Starshenko, {\it Mod. Phys. Lett.}
{\bf A10} (1995) 235.
\bibitem{SEK} M.A. Samuel, J. Ellis and M. Karliner, {\it 
Phys.Rev. Lett.} {\bf 74} (1995) 4380.
\bibitem{CKS} K.G. Chetyrkin, B.A. Kniehl and A. Sirlin, 
{\it Phys. Lett.} {\bf B402} (1997) 359.
\bibitem{HPade} F.Chishtie, V. Elias and T.G. Steele, 
{\it Phys. Rev.} {\bf D59} (1999) 105013 (hep-ph/9812498).
\bibitem{EJJKS} J. Ellis, I. Jack, D.R.T. Jones, M. Karliner 
and M.A. Samuel, {\it Phys. Rev.} {\bf D57} (1998) 2665. 
\bibitem{SVZ} M. A. Shifman, A.I. Vainshtein and V.I. Zakharov,  
{\it Nucl. Phys.} {\bf B147} (1979) 385.
\bibitem{ChKrT} K.G. Chetyrkin, N.V. Krasnikov and A.N. Tavkhelidze,
{\it Phys. Lett.} {\bf B76} (1978) 83;\\
A.L. Kataev, N.V. Krasnikov and A.A. Pivovarov, {\it Phys. Lett.}
{\bf B107} (1981) 115;\\
S.G. Gorishny, A.L. Kataev and S.A. Larin, {\it Phys. Lett.} 
{\bf B135} (1984) 457;\\
C.A. Dominquez and E. de Rafael, {\it Ann. Phys.} {\bf 174} (1987) 372.
\bibitem{BGen}
D.J. Broadhurst and S.C. Generalis, Preprint OUT-4102-8 (1982) (unpublished).
\bibitem{r0} F.J. Yndurain, Talk at the Conference in Warsaw, 1980 
(private communication)
\bibitem{r1} M.R. Pennington and G. Ross, {\it Phys. Lett.} {\bf B102}
(1981) 167.
\bibitem{r2} N.V. Krasnikov and A.A. Pivovarov, {\it Phys. Lett.} {\bf 
B116} (1982) 168.
\bibitem{r3} A.V. Radyushkin, Dubna Preprint E2-82-159 (1982), 
{\it JINR Rapid Comm.} N4 [78]-96 (1996) 9 (hep-ph/9907228)
\bibitem{r4} M.R. Pennington, R.G. Roberts and G.G. Ross, 
{\it Nucl. Phys.} {\bf B242} (1984) 69. 
\bibitem{r5} A.A. Pivovarov, {\it Sov.J. Nucl. Phys.} {\bf 54}  (1991) 676; 
{\it Z. Phys.} {\bf C53} (1992) 461.
\bibitem{r6} F. Le Diberder and A. Pich, {\it Phys. Lett.} {\bf B286} (1992) 147.
\bibitem{r7} M. Neubert, {\it Nucl. Phys.} {\bf B463} (1996) 511.
\bibitem{r8} M. Girone and M. Neubert, {\it Phys. Rev. Lett.} 76 (1996) 3061.
\bibitem{r9} C.J. Maxwell and D.G. Tonge, {\it Nucl. Phys.} {\bf B535} (1998) 19.
\bibitem{r10} B.V. Geshkenbein and B.L. Ioffe, {\it JETP Lett.} {\bf 70} (1999) 161.
\bibitem{r11} D.V. Shirkov, Preprint JINR-E2-2000-46 (hep-ph/0003242)
\bibitem{r13} 
G.`t Hooft in : Deeper Pathways in high energy physics, Proc. Orbis
Scientiae (Coral Gables, Florida 1977) eds., A. Perlmutter and L.F. Scott,
(Plenum , New York, 1977);\\ 
E. de Rafael in Lecture Notes in Physics, vol. 118,
eds. J.J. Alfonso and R. Tarrach (Springer, Berlin 1980) (1980).
\bibitem{Becchi}
C. Becchi, S. Narison, E. de Rafael and F.J. Yndurain, 
{\it Z. Phys.} {\bf C8} (1981) 335.
\bibitem{r12} J.G. K\"orner, F. Krajewski and A.A. Pivovarov, 
Preprint MZ-TH/99-66 (hep-ph/0003165) 
\bibitem{pmp} A.\ Palanques--Mestre and P.\ Pascual,
{\it Comm.\ Math.\ Phys.}\ {\bf 95} (1984) 277.
\bibitem{MB} M. Beneke, {\it Nucl. Phys.} {\bf B405} (1993) 424.  
\bibitem{JL}  R. Jost and J. Luttinger, {\it Helv.Phys.Acta} {\bf 23} (1950) 201.
\bibitem{BGK} D.J.  Broadhurst, J.A. Gracey and D. Kreimer,
{\it Z. Phys.} {\bf C75} (1997) 559.
\bibitem{ChT} F.V. Tkachov, {\it Phys. Lett.} {\bf B100} (1981) 65;\\
K.G. Chetyrkin and F.V. Tkachov, {\it Nucl. Phys.} {\bf B192} (1981) 159.
\bibitem{V} S.J. Brodsky, M. Melles and J. Rathsman, {\it Phys. Rev.} 
{\bf D60} (1999) 096006 (hep-ph/9906324)
\bibitem{JG1} J.A. Gracey, {\it Phys. Lett.} {\bf B317} (1993) 415. 
\bibitem{JG2} M. Ciuchini, S.E. Derkachev, J.A. Gracey and A.N. Manashov, 
{\it Phys. Lett.} {\bf B458} (1999) 117. 
\bibitem{NSVZ} V.A. Novikov, M.A. Shifman, A.I. Vainshtein and V. I. 
Zakharov, {\it Nucl. Phys.} {\bf B191} (1981) 301. 
\bibitem{KrPiv3} N.V. Krasnikov and A.A. Pivovarov, 
{\it Mod. Phys. Lett.} {\bf A11} (1996) 835.
\bibitem{CDPS} K.G. Chetyrkin, C.A. Dominguez, D. Pirjol and K. Schilcher, 
{\it Phys. Rev.} {\bf D51} (1995) 5090.
\bibitem{JM}
M. Jamin and M. M\"unz, {\it Z. Phys.} {\bf C66} (1995) 633.
\bibitem{GK} E. Gardi and M. Karliner, {\it Nucl. Phys.} {\bf B529} (1998) 383. 
\bibitem{Grunberg1} G. Grunberg,  {\it Phys. Rev.} {\bf D29} (1984) 2315.
\bibitem{Kr} N.V. Krasnikov, {\it Nucl. Phys.} {\bf B192} (1981) 497;\\
A.L. Kataev, N.V. Krasnikov and A.A. Pivovarov,{\it Nucl. Phys.} {\bf B198} (1982) 508.
\bibitem{SIPT} A. Dhar and V. Gupta, {\it Phys. Rev.} {\bf D29} (1984) 2822:\\
 V. Gupta, D.V. Shirkov  and O.V. Tarasov, {\it Int. J. Mod. Phys.} {\bf A6} (1991) 3381.
\bibitem{BL} S.J. Brodsky and H.J. Lu, {\it Phys. Rev.} {\bf D51} (1995) 3652.
\bibitem{PMS}
P.M. Stevenson, {\it Phys. Rev.} {\bf D23} (1981) 2916.
\bibitem{SM}
S.V. Mikhailov, {\it Phys. Lett.} {\bf B431} (1998) 387;\\
S.V. Mikhailov, {\it Phys. Rev.} {\bf D62} (2000) 034002 (hep-ph/9910389).
\bibitem{KS2}
A.L. Kataev and V.V. Starshenko, {\it Phys. Rev.} {\bf D52} (1995) 402.
\bibitem{r19} M. Beneke, V.M. Braun and N. Kivel, {\it Phys. Lett.} {\bf B404} 
(1997) 315.
\bibitem{r20} C.J. Maxwell, {\it Phys. Lett.} {\bf B409} (1997) 382.
\bibitem{r22} C.N. Lovett-Turner and C.J. Maxwell, {\it Nucl. Phys.} {\bf B452} (1995) 188.
\bibitem{r23} M. Beneke (private communications).
\bibitem{r18} G. Grunberg, {\it Phys. Lett.} {\bf B153} (1985) 427.
\bibitem{r24} P. Ball, M. Beneke and V.M. Braun, {\it Nucl. Phys.} 
{\bf B452} (1995) 563.
\bibitem{Rid} G. Altarelli, P. Nason and G. Ridolfi, {\it Z. Phys.}{\bf C68} 
(1995) 257.
\bibitem{ChKuP} K.G. Chetyrkin, J.H. K\"uhn and A.A. Pivovarov, 
{\it Nucl. Phys.} {\bf B533} (1998) 473:\\
A. Pich and J. Prades, {\it JHEP} 9806:013 (1998), {\it ibid.} 9910:004 (1999)
\bibitem{Q2}
D.V. Shirkov and I.L. Solovtsov, {\it Phys. Rev. Lett.} {\bf 79} (1997) 1209;\\
K.G. Chetyrkin, S. Narison and V. I. Zakharov,
{\it Nucl. Phys.} {\bf B550} (1999) 353:\\
M.N. Chernodub, F.V. Gubarev, M.I. Polikarpov and V.I. Zakharov, 
Preprint ITEP-TH-74-99 (hep-ph/0003006). 
\bibitem{Yndc} F.J. Yndurain, {\it Nucl. Phys.} {\bf B517} (1998) 324 
(hep-ph/9708300)
\bibitem{95}
R. Harlander and M. Steinhauser, {\it Phys. Rev.} {\bf D56} (1997) 3980.
\bibitem{96}
D.J. Broadhurst, J. Fleicher and O.V. Tarasov, {\it Z. Phys.} 
{\bf C60} (1993) 287.
\bibitem{97}
P.A. Baikov and D.J. Broadhurts, Proceedings of the IV Int. Workshop 
on Software Engeneering and Artificial Intelligence for High 
Energy and Nuclear Physics, Pisa, Italy 3-8 April, 1995, World Scientific, 1995, Ed. by B. Denby and D. Perret-Gallix, p.167 (hep-ph/9504398).
\bibitem{98} 
K.G. Chetyrkin, J. K\"uhn and M. Steinhauser, {\it Nucl. Phys.} 
{\bf B505} (1997) 40.
\bibitem{99}
A.H. Hoang and T. Teubner, {\it Nucl. Phys.} {\bf B519} (1999) 285.
\bibitem{100}
S. Peris and E. de Rafael, {\it Nucl. Phys.} {\bf B500} (1997) 325.

\end{thebibliography}
\end{document}